\documentclass[fleqn,usenatbib]{mnras}

\usepackage{newtxtext,newtxmath}

\usepackage[T1]{fontenc}

\DeclareRobustCommand{\VAN}[3]{#2}
\let\VANthebibliography\thebibliography
\def\thebibliography{\DeclareRobustCommand{\VAN}[3]{##3}\VANthebibliography}

\usepackage{graphicx}	
\usepackage{amsmath}	
\usepackage{subcaption}
\usepackage{xcolor}
\usepackage{xspace}

\newcommand{\Mch}{\ensuremath{{\rm{M}}_{\rm{ch}}}\xspace}

\title[Double detonations: core and He shell masses]{Double detonations: variations in Type Ia supernovae due to different core and He shell masses -- II: synthetic observables}

\author[C. E. Collins et al.]{
Christine E. Collins,$^{1,2}$\thanks{E-mail: c.collins@gsi.de}
Sabrina~Gronow,$^{3,4}$
Stuart A. Sim$^{2}$
and Friedrich~K.~R\"{o}pke$^{4,5}$
\\
$^{1}$GSI Helmholtzzentrum f\"{u}r Schwerionenforschung, Planckstraße 1, 64291 Darmstadt, Germany\\
$^{2}$Astrophysics Research Center, School of Mathematics and Physics, Queen's
University Belfast, Belfast BT7 1NN, Northern Ireland, UK\\
$^{3}$Zentrum f\"ur Astronomie der Universit\"at Heidelberg,
Astronomisches Rechen-Institut, M\"{o}nchhofstr. 12-14, 69120 Heidelberg, Germany\\
$^{4}$Heidelberger Institut f\"{u}r Theoretische Studien,
Schloss-Wolfsbrunnenweg 35, 69118 Heidelberg, Germany\\
$^{5}$Zentrum f\"ur Astronomie der Universit\"at Heidelberg, Institut f\"ur
theoretische Astrophysik, Philosophenweg 12, 69120 Heidelberg, Germany\\
}

\date{Accepted XXX. Received YYY; in original form ZZZ}

\pubyear{2022}

\begin{document}
\label{firstpage}
\pagerange{\pageref{firstpage}--\pageref{lastpage}}
\maketitle

\begin{abstract}
Double detonations of sub-Chandrasekhar mass white dwarfs are a promising explosion scenario
for Type Ia supernovae,
whereby a detonation
in a surface helium shell triggers a secondary detonation
in a carbon-oxygen core.
Recent work has shown that low mass helium shell models reproduce observations of normal SNe Ia.
We present 3D radiative transfer simulations for {a suite of} 3D simulations
of the double detonation explosion scenario for a range
of shell and core masses.
We find light curves broadly able to reproduce the faint end of the width-luminosity relation shown by SNe Ia,
however, we find that all of our models show extremely red colours,
not observed in normal SNe Ia.
This includes our lowest mass helium shell model.
We find clear \ion{Ti}{II} absorption features in the model spectra, which would lead to classification as peculiar SNe Ia,
as well as line blanketing in some lines of sight by singly ionised Cr and Fe-peak elements.
Our {radiative transfer simulations show that these explosion models} remain promising to explain peculiar SNe Ia.
Future full non-LTE simulations may improve the agreement of these {explosion} models with observations of normal SNe Ia.
\end{abstract}




\section{Introduction}

Type Ia supernovae (SNe Ia) are extremely well studied,
predominantly due to their use as distance indicators in cosmology.
Despite this, we still do not know the exact progenitor or explosion
mechanism of SNe Ia (see e.g. \ \citealt{maoz2014b} for a review).
It is understood that SNe Ia are the thermonuclear explosion of a white dwarf (WD),
but it remains an open question whether these explode as the WD
nears the Chandrasekhar limit, or whether sub-Chandrasekhar mass
WDs are responsible for SNe Ia.
There is growing evidence suggesting that sub-\Mch WDs account for at least some population of SNe Ia
\citep[e.g.][]{scalzo2014a, blondin2017a, goldstein2018a, polin2019a, bulla2020a}.
Studies of pure detonations of sub-\Mch \ carbon-oxygen WDs
(which did not consider a physical ignition mechanism) have shown reasonable agreement
with observations of SNe Ia \citep{sim2010a, blondin2017a, shen2018b, shen2021a},
such as reproducing the observed width-luminosity relation \citep{phillips1993a}.

A widely discussed explosion mechanism for sub-\Mch WDs is the double
detonation \citep[see eg.][]{taam1980a, nomoto1980a, nomoto1982a, livne1990a, woosley1994b,
hoeflich1996a, nugent1997a}.
In this scenario, a helium detonation is ignited in a helium shell on a carbon-oxygen white dwarf. 
The helium detonation then ignites a secondary carbon detonation in the core.
However, in such early models, relatively massive helium shells were considered, which produced light curves
and spectra inconsistent with observations.
There has been renewed interest in the double detonation,
due to considering lower helium shell masses \citep{bildsten2007a, shen2009a, fink2010a, shen2010a}
leading to reduced discrepancies with observations
 \citep{kromer2010a, woosley2011b}.
Nevertheless, discrepancies with
observations still remained, such as red colours due to absorption
by the products of the helium shell detonation 
\citep{kromer2010a,
sim2012a, polin2019a, gronow2020a, shen2021b}.
However, \cite{townsley2019a} found good agreement
to observations of the normal SN 2011fe for their model
considering a minimal helium shell mass.
Similarly \cite{shen2021b} found that their minimal He shell mass
models were able to reproduce normal SNe Ia.

{Double detonation models have been suggested to explain
a number of peculiar SNe Ia \citep{inserra2015a, Jiang2017a, de2019a, dong2022a},
which showed unusually red colours, early flux excesses and \ion{Ti}{II} absorption features.
The red colours can be explained by absorption at blue wavelengths by the
products of the helium shell detonation, as can the \ion{Ti}{II} absorption features \citep[e.g.][]{kromer2010a, polin2019a}, since Ti is predicted to be synthesised in the high velocity outer
layers of the ejecta \citep{fink2010a}.
The shell detonation can also explain the early flux excess, since double detonation
models predict surface radioactive material, synthesised in the helium shell detonation, which may lead to an early bump in the light curve \citep[e.g.][]{noebauer2017a}.
}

\cite{gronow2021a} carried out 3D double detonation explosion simulations,
varying the masses of the core WD and helium shell.
They presented the predicted bolometric light curves for
each of these models, and made comparisons with observations.
They found that the bolometric light curves showed a strong angle dependence,
and appeared to be too asymmetric compared to the bolometric dataset
constructed by \cite{scalzo2019a} from well-observed SNe Ia.
In this paper, we carry on from this work and present
band-limited light curves and spectra from 3D radiative transfer simulations
for the models of \cite{gronow2021a}.

\section{Methods}

\subsection{Radiative transfer}

To produce light curves and spectra
for the models presented by \cite{gronow2021a},
we carried out 3D radiative transfer simulations
using
\textsc{artis},
a time-dependent multi-dimensional Monte Carlo radiative
transfer code, developed by \cite{sim2007b} and \cite{kromer2009a},
based on the methods
of \cite{lucy2002a, lucy2003a, lucy2005a}.
\cite{shingles2020a} have added full non-LTE and non-thermal capabilities to \textsc{artis},
however in this work we do not make use of these,
and use a non-LTE approximation as described by \cite{kromer2009a}.

The explosion model densities and 
nucleosynthetic abundances (see \citealt{gronow2021a})
were mapped to a $50^3$ Cartesian grid.
We assume the models to be in homologous expansion.
In each simulation, 3.36$\times 10^7$ Monte Carlo
energy packets were propagated through the explosion ejecta,
between 0.1 to 100 days after explosion, using 110
logarithmically spaced time steps.
Escaping packets of photons were binned into 100
equal solid-angle bins, defined by spherical polar coordinates
relative to the positive $z$-axis.
Additionally, we use `virtual' packets as described by 
\cite{bulla2015a} to obtain detailed line of sight
spectra in specific lines of sight.
We use the atomic data set compiled by \cite{gall2012a},
sourced from \cite{kurucz2006a}.

\subsection{Models}

\begin{figure*}
\centering
    \begin{subfigure}[b]{0.95\textwidth}
        \caption{M08\_03}
        \includegraphics[width=0.95\textwidth]{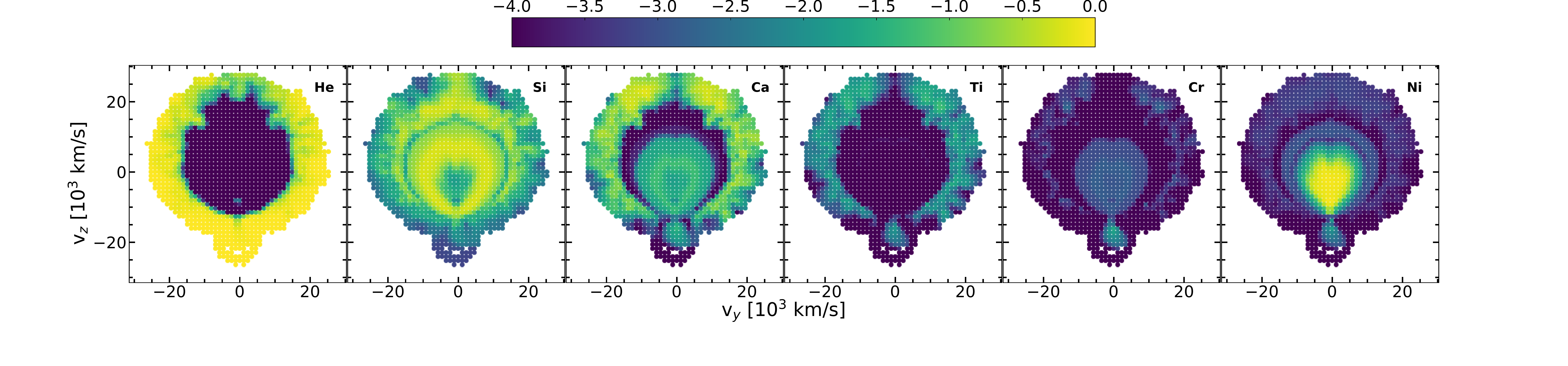}
    \end{subfigure}

    \begin{subfigure}[b]{0.95\textwidth}
        \caption{M10\_03}
                \includegraphics[width=0.95\textwidth]{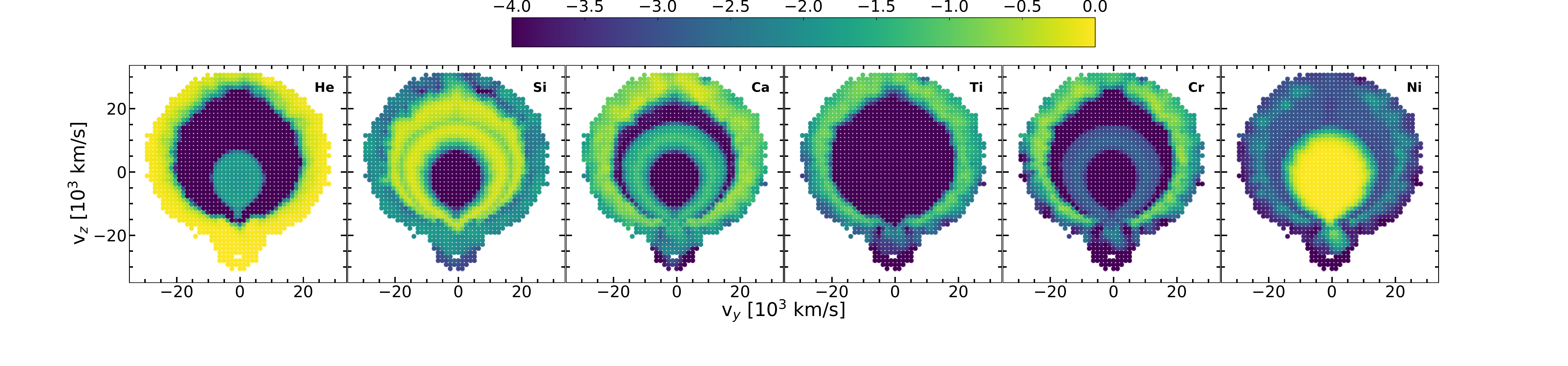}
    \end{subfigure}
    
    \begin{subfigure}[b]{0.95\textwidth}
        \caption{M10\_05}
        \includegraphics[width=0.95\textwidth]{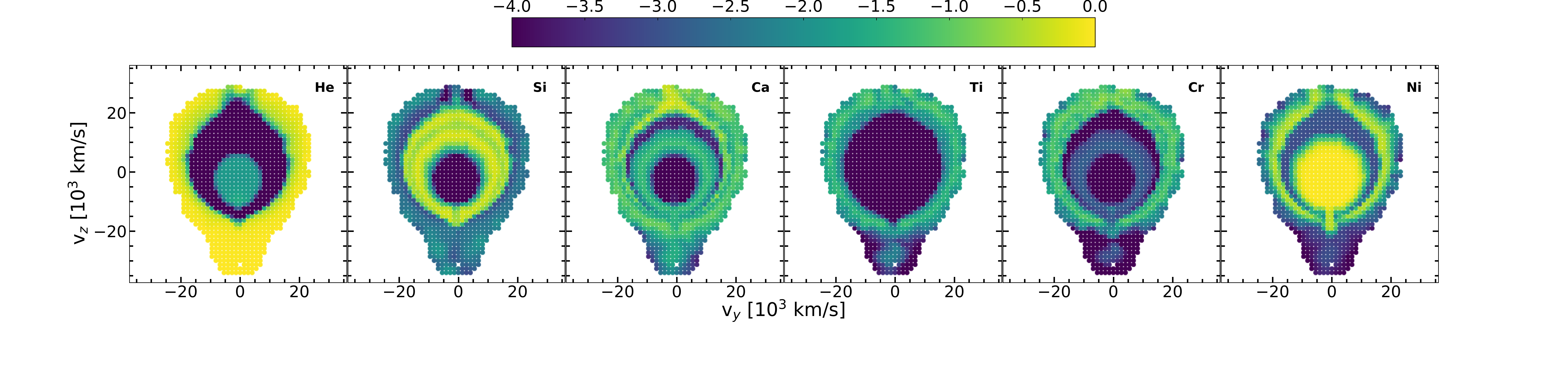}
    \end{subfigure}
    
    \begin{subfigure}[b]{0.95\textwidth}
        \caption{M10\_10}
        \includegraphics[width=0.95\textwidth]{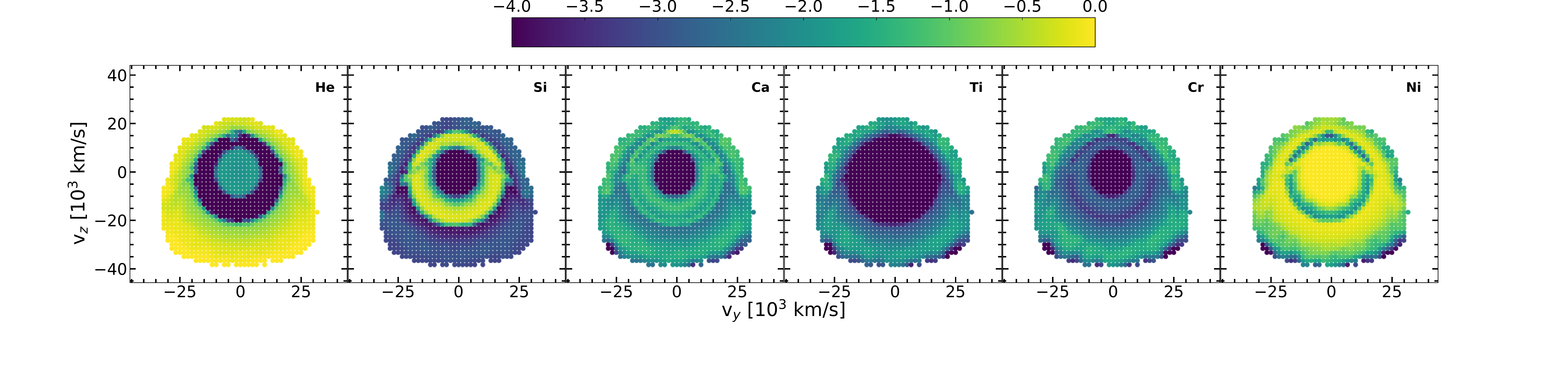}
    \end{subfigure}
    \caption{Colour scale indicates the abundances in log mass fraction of key species for selected models. 
    Included are models showing
    a secondary detonation by the converging shock (M08\_03),
    scissors (M10\_05) and edge-lit (M10\_10) mechanisms. 
    }
    \label{fig:composition}
\end{figure*}

\begin{table*}
    \caption{Model parameters. Model masses listed are pre-relaxation (see \citet{gronow2021a} for details). We keep the model names from \citet{gronow2021a}.
    We include the abundances of $^{56}$Ni synthesised in each model since this 
    predominantly determines the model light curve brightness. 
    For other nucleosynthetic abundances see \citet{gronow2021a}.
        We show the ignition mechanism of the secondary detonation, which are the converging shock mechanism `cs'; the scissors mechanism `s'; the converging shock
    mechanism, but with carbon burning at the convergence of the helium detonation,
    similar to the scissors mechanism `(s,)cs';
    and the edge-lit mechanism `edge'. We note that in Model M10\_02 the secondary detonation due to the converging shock was artificially ignited by hand.
    Also shown are the angle-averaged peak light curve brightnesses, as well as
    time of peak, and the decline rate in B band over 15 days
    from peak, $\Delta$m$_{15}$(B).}
    \label{tab:paramstudy-observables}
    \centering
    \resizebox{\textwidth}{!}{\begin{tabular}{llllllllllll}
        \hline
        & M08\_03 & M08\_05 & M08\_10\_r & M09\_03 & M09\_05 & M09\_10\_r & M10\_02 & M10\_03 & M10\_05 & M10\_10 & M11\_05\\ \hline
        Core Mass [M$_\odot$] & 0.803 & 0.803 & 0.795 & 0.905 & 0.899 & 0.888 & 1.005 & 1.028 & 1.002 & 1.015 & 1.100 \\
        Shell Mass [M$_\odot$] & 0.028 & 0.053 & 0.109 & 0.026 & 0.053 & 0.108 & 0.020 & 0.027 & 0.052 & 0.090 & 0.054 \\ \hline
        $^{56}$Ni Core [M$_\odot$] & $1.3\times10^{-1}$ & $2.0\times10^{-1}$ & $3.1\times10^{-1}$ & $3.3\times10^{-1}$ & $3.8\times10^{-1}$ & $4.8\times10^{-1}$ & $5.4\times10^{-1}$ & $5.9\times10^{-1}$ & $5.4\times10^{-1}$ & $7.2\times10^{-1}$ & $8.3\times10^{-1}$ \\
        $^{56}$Ni Shell [M$_\odot$] & $9.9\times10^{-7}$ & $6.7\times10^{-5}$ & $1.5\times10^{-2}$ & $1.0\times10^{-6}$ & $2.0\times10^{-3}$ & $2.6\times10^{-2}$ & $1.9\times10^{-6}$ & $6.0\times10^{-5}$ & $8.2\times10^{-3}$ & $3.9\times10^{-2}$ & $1.2\times10^{-2}$ \\\hline
        Mechanism                       & cs      & (s,) cs & s       & (s,) cs & (s,) cs & s       & art cs  & (s,) cs & s       & edge & edge \\ \hline
        M$_{\mathrm{bol, max}}$         & -17.6   & -17.9   & -18.4   & -18.4   & -18.5   & -18.8   & -18.9   & -19.0   & -18.9   & -19.2 & -19.3  \\
        M$_{\mathrm{U, max}}$           & -16.3   & -16.7   & -17.5   & -17.8   & -17.9   & -18.7   & -18.9   & -19.0   & -18.9   & -19.2 & -19.4  \\
        M$_{\mathrm{B, max}}$           & -17.0   & -17.2   & -18.0   & -18.3   & -18.3   & -18.9   & -19.0   & -19.2   & -19.0   & -19.4 & -19.6   \\
        M$_{\mathrm{V, max}}$           & -18.3   & -18.6   & -19.2   & -19.3   & -19.4   & -19.7   & -19.7   & -19.8   & -19.7   & -20.1 & -20.1  \\
        M$_{\mathrm{R, max}}$           & -18.4   & -18.8   & -19.2   & -19.1   & -19.3   & -19.4   & -19.4   & -19.5   & -19.4   & -19.6  & -19.6 \\ \hline
        t$_{\mathrm{bol, max}}$ [days] & 17.4    & 18.1    & 17.9    & 18.1    & 18.1    & 17.2    & 17.3    & 17.1    & 17.4    & 16.6 & 16.1   \\
        t$_{\mathrm{U, max}}$ [days]    & 15.1    & 16.4    & 16.4    & 15.7    & 16.7    & 15.6    & 15.1    & 15.2    & 16.0    & 14.3 & 13.8    \\
        t$_{\mathrm{B, max}}$ [days]    & 17.5    & 17.8    & 17.8    & 18.0    & 18.1    & 17.2    & 17.2    & 16.9    & 17.4    & 16.3 & 15.6    \\
        t$_{\mathrm{V, max}}$ [days]    & 19.2    & 19.9    & 19.5    & 19.9    & 19.9    & 19.0    & 19.6    & 19.4    & 19.5    & 18.7  & 18.6  \\
        t$_{\mathrm{R, max}}$ [days]   & 18.3    & 19.2    & 17.5    & 18.0    & 18.0    & 16.7    & 17.6    & 17.4    & 17.2    & 16.7   & 16.7 \\ \hline
        $\Delta \mathrm{m} _{15}$(B)        & 1.38    & 1.54    & 1.94    & 1.52    & 1.96    & 2.01    & 1.51    & 1.73    & 1.94    & 1.42 & 1.35   \\ \hline
        
    \end{tabular}}
\end{table*}

We carry out radiative transfer simulations for the 3D hydrodynamical explosion simulations by \cite{gronow2021a}.
They investigated 11 models with a range of different
core and shell masses.
The model core masses range from 0.8~M$_\odot$ \ to 1.1~M$_\odot$ \ and the shell masses range from
0.02~M$_\odot$ \ to 0.1~M$_\odot$.
The shell and core mass combinations were chosen to match models in previous work
\citep[eg.][]{woosley2011b,polin2019a,townsley2019a}.
A core mass of $\sim$ 1 M$_\odot$ \ has been found to produce models of similar brightness
to normal SNe Ia \citep[eg.][]{sim2010a, kromer2010a}.
Given that the initial model masses vary significantly between models,
parameters important for determining the evolution of the explosion,
such as the central density, also vary between models.
In all models, a helium detonation is ignited at a point on the positive $z$-axis
in the helium shell, as was described by \cite{gronow2021a}.
Three different mechanisms were found to ignite the secondary core detonation,
and one model did not dynamically ignite a secondary detonation.
The three mechanisms were the converging shock \citep[eg.][]{livne1990a, fink2007a, shen2014a},
edge-lit \citep[eg.][]{livne1990b} and scissors
mechanism \citep{forcada2006a, gronow2020a, gronow2021a}.
The explosion mechanism for each model is listed in Table~\ref{tab:paramstudy-observables}.
Following \cite{gronow2021a}, the models are named indicating the
original masses of the core and shell, which are listed in Table~\ref{tab:paramstudy-observables}.
Figure \ref{fig:composition} indicates the relative abundances 
of key species in the ejecta for models showing each of the three
detonation mechanisms found in this study.
All three mechanisms produce highly asymmetrical
ejecta, as has previously been discussed by
\cite{gronow2021a}.

\section{Results}

We have computed the light curves and spectra for
each of the explosion models.
We first present the light curves
in Section~\ref{sec:lightcurves},
and then present the spectra in Section~\ref{sec:param-study-spectra}.

\subsection{Light curves}
\label{sec:lightcurves}

In this section, we discuss the light curves predicted 
for the models presented by \cite{gronow2021a}.
We first present the angle-averaged properties of the
light curves in Section~\ref{sec:lightcurves-angleave-bandlimited-paramstudy} and their colour evolution in Section~\ref{sec:colourevolution-paramstudy-angleaveraged}.
We then discuss the viewing-angle 
dependence of the light curves in Section~\ref{sec:viewingangle-lightcurves}.
We compare the light curves to the normal SN~2011fe {\citep{nugent2011a}},
the over-luminous SN~1991T {\citep{filippenko1992a}} and the sub-luminous 
SN~1999by {\citep{garnavich2004a}} to determine the ability of these models
to reproduce the main classes of observed SNe~Ia,
and also compare to the sample of observed SNe~Ia
from \cite{hicken2009b}.

\subsubsection{Angle-averaged light curves}
\label{sec:lightcurves-angleave-bandlimited-paramstudy}

\begin{figure*}
    \begin{center}
        \includegraphics[width=0.8\textwidth]{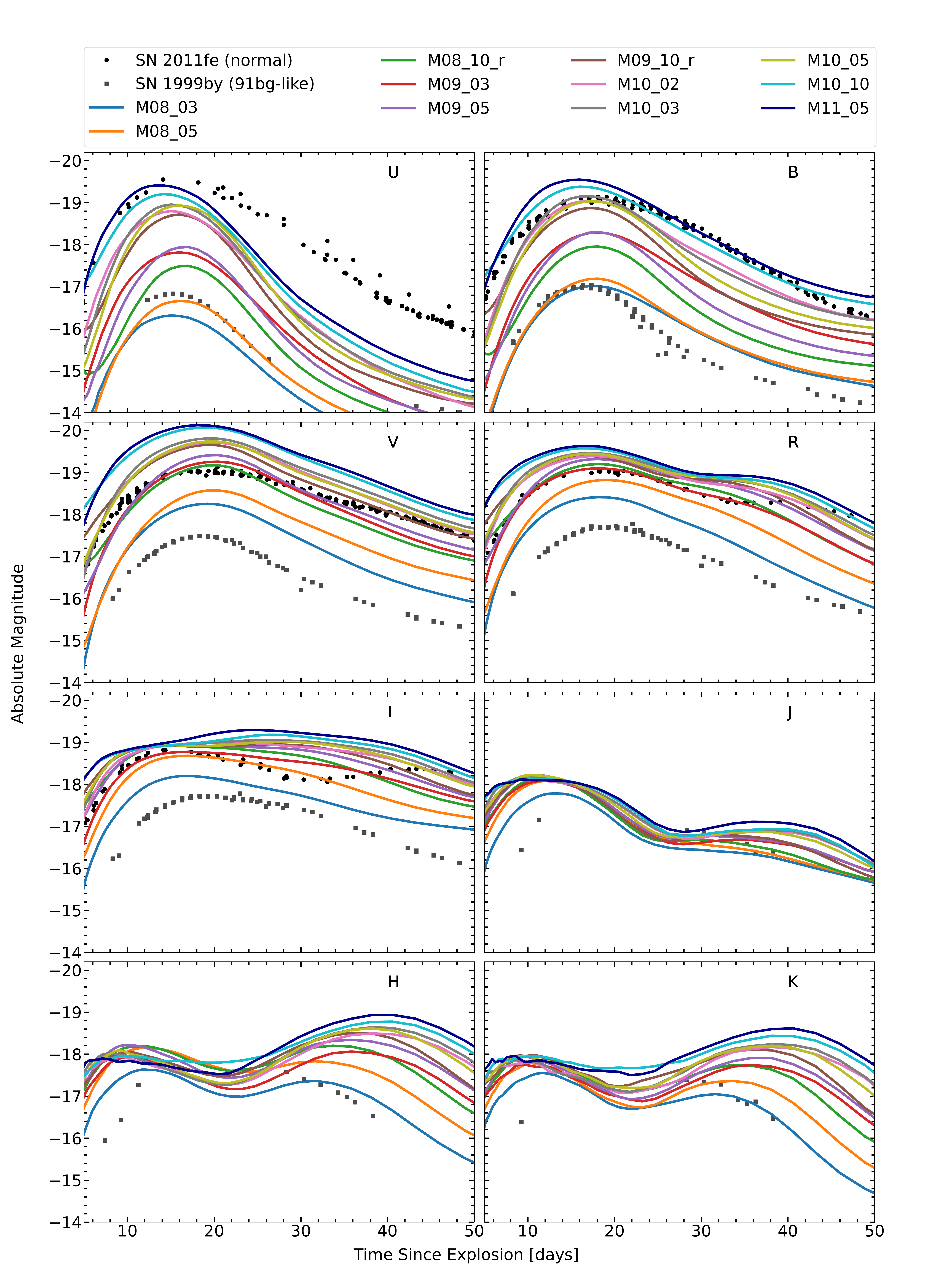}
    \end{center}
    \caption{Angle-averaged band limited light curves for our parameter
    study models. Also plotted are the light curves of the normal SN~2011fe and
    91bg-like SN~1999by.}
    \label{fig:lightcurvesUBVRIJHK-paramstudy}
\end{figure*}

To make general comparisons between the models, 
we first discuss the angle-averaged light curves.
As indicated by the asymmetries in the explosion 
ejecta (see Figure~\ref{fig:composition}),
the angle-averaged light curves do not give a complete representation
of the models as a whole.
We discuss the line of sight dependent light curves in Section~\ref{sec:viewingangle-lightcurves}.

In Figure~\ref{fig:lightcurvesUBVRIJHK-paramstudy} we show the band limited
angle-averaged light curves for each of the models.
We also show the light curves of the normal SN~2011fe, 
and the 91bg-like
SN~1999by to indicate the fainter end of observed SNe Ia.
As anticipated from the range of model masses,
specifically the range in masses of $^{56}$Ni synthesised
in the models (see Table~\ref{tab:paramstudy-observables}),
the light curves show a wide range of brightnesses.
These span from fainter, 91bg-like brightnesses
to normal brightness SNe Ia, as previously discussed
by \cite{gronow2021a} for the model bolometric light curves.
Interestingly, the increase in model mass from M10\_10 
to M11\_05 does not show a significant increase in light curve 
brightness.
This may indicate that this scenario does not explain the brightest observed SNe Ia. 
Since increasing model mass increases the central density, leading to higher abundances
of Fe-group material and lower abundances of IME's,
increasing the model mass further to increase the 
brightness will lead to an over-production of Fe-group elements.

In this work, we do not present the early light curve evolution (< 5 days), since very early times are computationally expensive. 
However, we note that particularly the massive He shell models
tend to show an excess of flux in the bluer bands at early times, similar to that found by \cite{noebauer2017a}, due to the surface radioactive material synthesised during the He shell detonation.

Compared to observations, the models tend to be too faint in the U band and too bright
in V band, such that the models are redder than observations.
We discuss the model colour evolution in Section~\ref{sec:colourevolution-paramstudy-angleaveraged}.
In agreement with SN~1999by, our faintest models (Models M08\_03 and M08\_05)
do not show secondary maxima in the R and I bands.
While our brightest models do show secondary maxima (in agreement with normal SNe Ia),
they do not match the times of the secondary maxima in SN~2011fe.
The NIR secondary maxima in SNe Ia
are attributed to the recombination of
doubly ionised iron group elements to singly ionised in the inner iron-rich
regions of the ejecta \citep{kasen2006b, kromer2009a, jack2015a}, as the ejecta expands and cools.
When this occurs, the flux from the UV and blue parts of the spectrum
is most effectively redistributed to the red and near-infrared by fluorescence,
causing the secondary maxima.
It is likely that the temperature in the ejecta is underestimated
in our simulations, since we do not
include a full non-LTE solution {(see discussion in Section~\ref{sec:colourevolution-paramstudy-angleaveraged})}.
Therefore it is plausible that this is the reason
that our secondary maxima in the R band occur too early.

\subsubsection{Angle-averaged colour evolution}
\label{sec:colourevolution-paramstudy-angleaveraged}

\begin{figure*}
    \centering
    \includegraphics[width=0.95\textwidth]{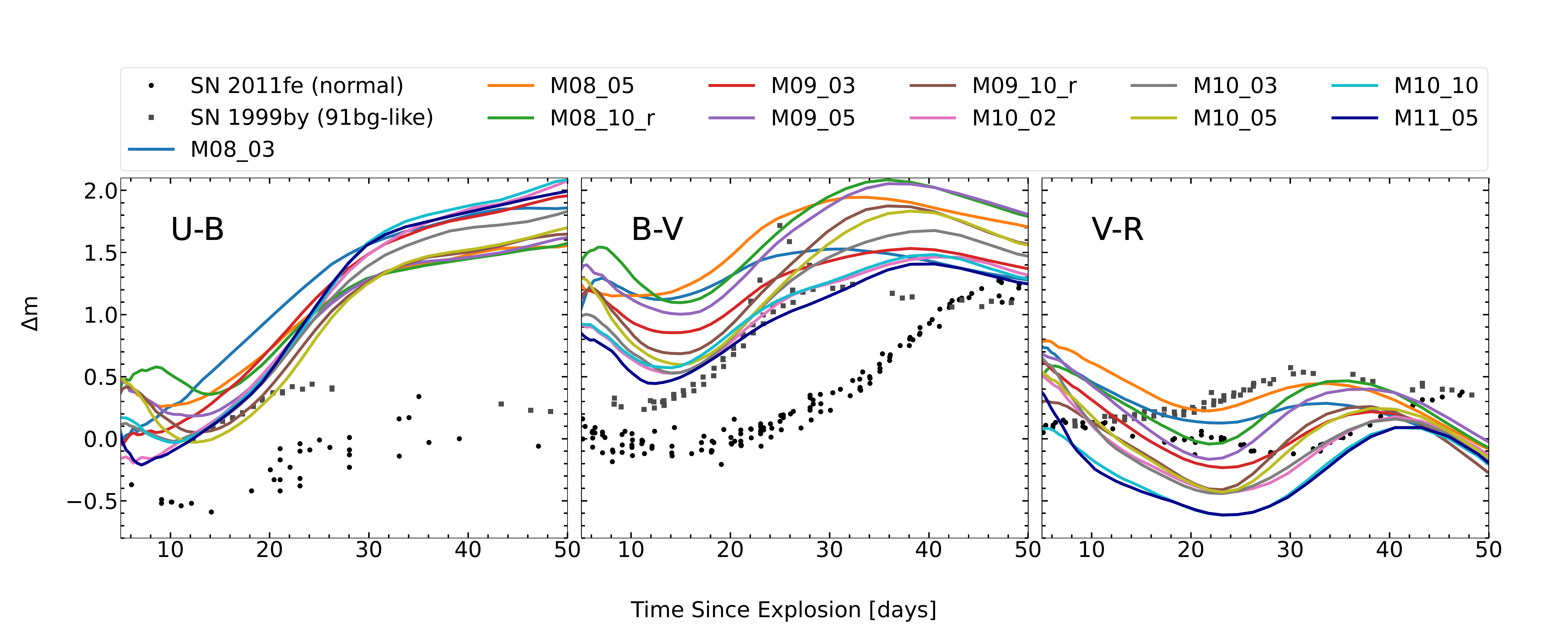}
    \caption{Angle-averaged U-B, B-V, and V-R colour evolution of our models. The colours of the normal SN~2011fe and
    91bg-like SN~1999by are also plotted.}
    \label{fig:colourevolution-paramstudy}
\end{figure*}

As previously discussed, 
many recent double detonation models show B$-$V
colours too red compared to observations 
\citep{kromer2010a, woosley2011b, polin2019a, gronow2020a, shen2021b}.
This is predominantly due
to line blanketing caused by heavy elements in
the outer layers of the ejecta, synthesised in the
helium detonation.
\cite{townsley2019a}, however, do not find these extremely red
colours for their minimal 
helium shell mass model, and \cite{shen2021b} also find
their minimal mass helium shell models do not show such red colours.

Our model shell masses range from a minimal shell mass of 
$0.02$ M$_\odot$ \ (Model M10\_02) to thicker shell masses
of $\sim 0.1$ M$_\odot$.
Compared to early double detonation models, this is still a low helium shell mass (e.g. \citealt{nugent1997a} considered a helium shell mass
of $0.2$ M$_\odot$).
In Figure~\ref{fig:colourevolution-paramstudy} we plot the 
angle-averaged colour evolution for each of the models.
We discuss colour in specific lines of sight
in Section~\ref{sec:viewingangle_B-V_colour}.

Generally, colour evolution is a temperature effect,
such that brighter, hotter SNe Ia show bluer
colours than fainter, cooler SNe Ia (e.g. \citealt{tripp1998a}).
We do find this general trend for our models, such that
brighter models tend to show bluer colours, however,
none of our angle-averaged light curves show B-V colours similar to
observed normal SNe Ia with similar peak brightness,
as the colours are extremely red.
Even our lowest shell mass model M10\_02 shows B-V
colours significantly redder than the normal SN~2011fe.
Therefore we do not confirm the results of
\cite{townsley2019a} and \cite{shen2021b} that a minimal helium shell mass
leads to normal SNe Ia colours.

Similarly, the U-B model colours do not reproduce
those expected for normal SNe Ia, due to line
blanketing of the spectra at these wavelengths, 
caused by the heavy elements synthesised in the helium shell
detonation.
The V-R colours are more similar to those of
normal SNe Ia than the bluer bands.
As indicated by the light curves, these bands are not
as strongly affected by the excess absorption
caused by the burning products of the helium shell detonation.
However, we still do not find the expected V-R colour
evolution.
This is likely in part due to the predicted timing of the R band
secondary maxima, as discussed in Section~\ref{sec:lightcurves-angleave-bandlimited-paramstudy}.

It may be that ionisation effects are contributing to the redness of the models,
as a result of the
assumptions made in the radiative transfer calculations in \textsc{artis}
(i.e.\ not solving the full non-LTE equations of statistical equilibrium).
Full non-LTE simulations of sub-\Mch, helium ignited models
were found to be very blue at maximum light in non-LTE simulations by \cite{hoeflich1996a} and \cite{nugent1997a}.
Similar results were found by \cite{blondin2017a}, who
found that their non-LTE simulations of sub-\Mch, pure detonation (i.e. no helium shell) models showed similar colours to observations,
and for fainter models, even bluer colours than similar
brightness \Mch models.
\cite{shen2021a} also found that their non-LTE, pure-detonation models were more highly ionised than LTE simulations {after maximum light}, and showed less \ion{Fe}{II} absorption.
{At B-band maximum light, the greatest difference in B-V colour of their non-LTE simulations was $\sim 0.25$ mag bluer than their LTE simulation.
However, at 15 days after maximum light, the maximum difference in B-V colour of the non-LTE simulation was $\sim 1$ mag bluer than the LTE simulation.
The maximum light B-V colour of our least red model (M11\_05)
is 0.5 mag, while SN~2011fe has a B-V $\approx$ 0 mag at maximum light.
}
Therefore, a more accurate, full non-LTE treatment of the radiative transfer
may reduce the apparent discrepancies between our double detonation model colours and observations{, but is unlikely to fully reconcile the discrepancy}.
{In addition, we note that} the full non-LTE simulation of a helium detonation model, with no secondary core detonation, presented by
\cite{dessart2015a} still showed red colours, despite the non-LTE treatment.

\subsubsection{Line of sight dependent light curves}
\label{sec:viewingangle-lightcurves}

\begin{figure*}
\centering
    \begin{subfigure}[b]{\textwidth}
        \caption{M08\_03}
        \includegraphics[width=\textwidth]{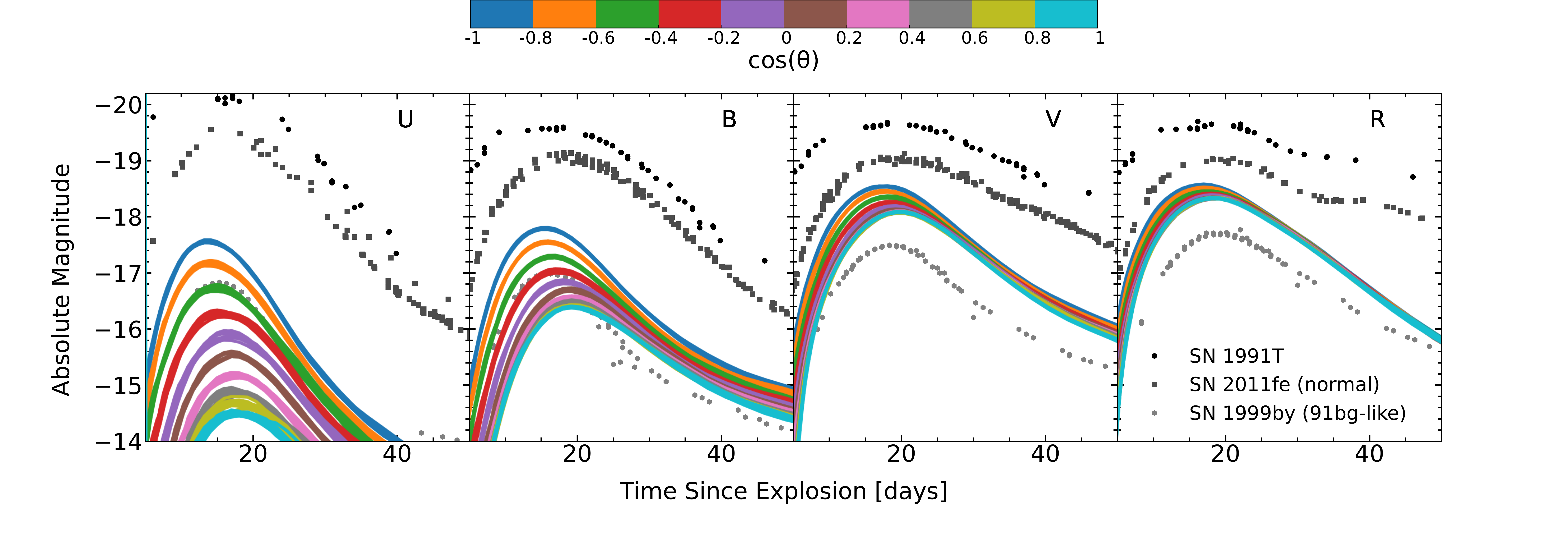}
    \end{subfigure}
    
    \begin{subfigure}[b]{\textwidth}
        \caption{M10\_05}
        \includegraphics[width=\textwidth]{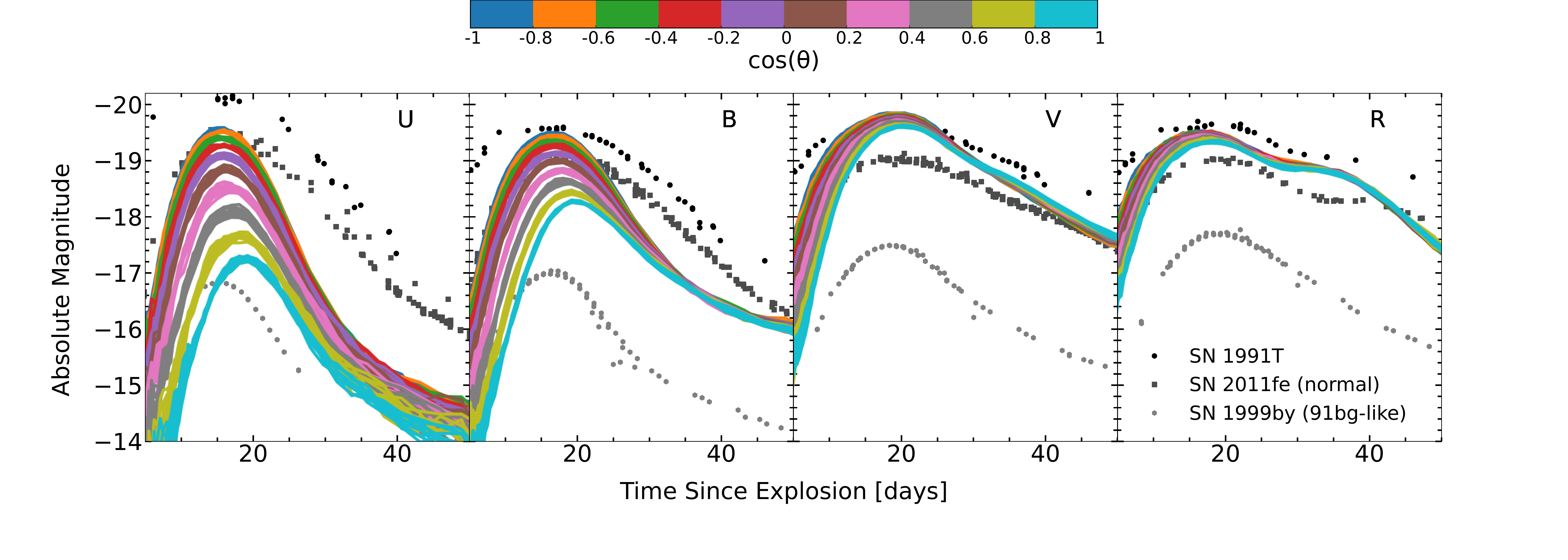}
    \end{subfigure}
    
    \begin{subfigure}[b]{\textwidth}
        \caption{M10\_10}
        \includegraphics[width=\textwidth]{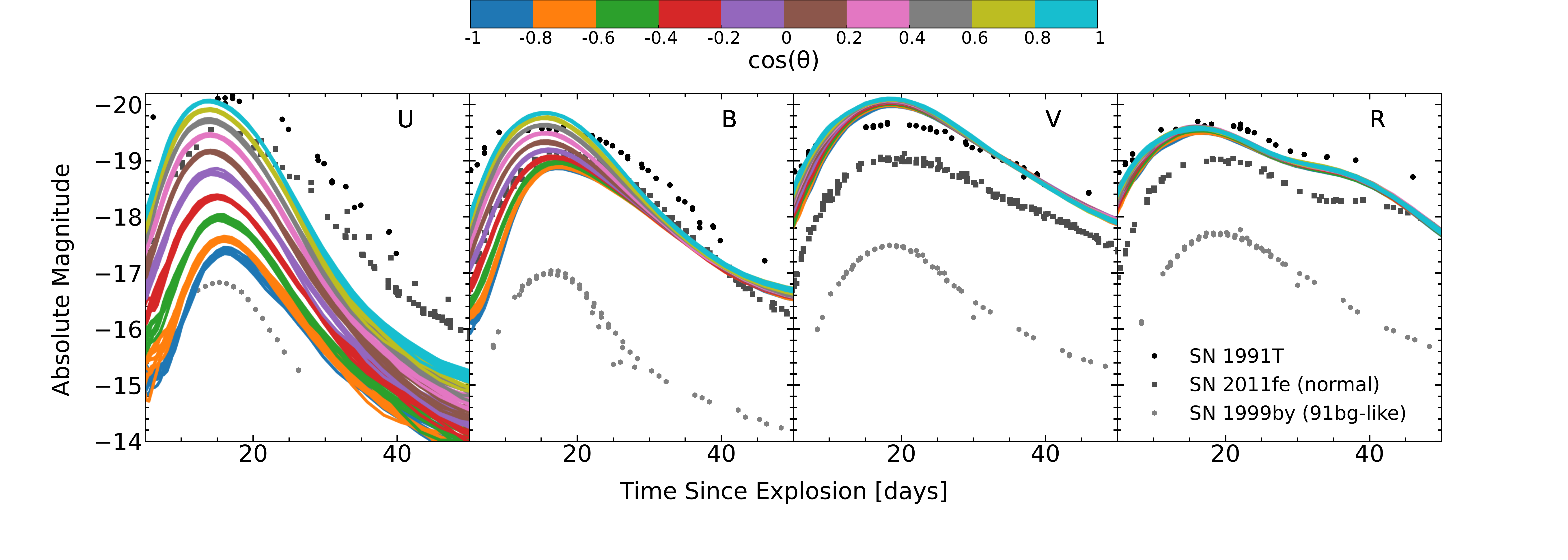}
    \end{subfigure}

    \caption{Light curves of Models M10\_10, M10\_05 and M08\_03,
    where escaping Monte Carlo packets (representing packets of photons) have been binned
    in 10 equal solid-angle bins in cos($\mathrm{\theta}$), indicated by the colour coding,
    and for each angle bin in cos($\mathrm{\theta}$) there are 10 equal solid-angle bins in $\mathrm{\phi}$. Since the models are close to asymmetric, we do not
    find significant variation in $\mathrm{\phi}$.
    The U and B bands show a strong angle dependence in cos($\mathrm{\theta}$), which becomes less significant with
    time as the ejecta become more optically thin.}
    \label{fig:lightcurvesM10_05}
\end{figure*}

In this section we discuss the light curves produced
in different lines of sight.
We show light curves for 100 viewing angles,
where escaping Monte Carlo packets are
divided into equal solid-angle bins (i.e. binned on a uniform grid in cos($\theta$) and $\phi$, where $\theta$ and $\phi$ are the usual spherical polar angles).

In Figure~\ref{fig:lightcurvesM10_05}, we plot the
viewing-angle dependent light curves
for Models M08\_03, M10\_05 and
M10\_10.
Each of these models showed a different mechanism to
ignite the secondary core detonation; converging shock, scissors mechanism and edge-lit, respectively.
The ejecta composition for these models is shown in 
Figure~\ref{fig:composition}.
The U and B band light curves show an extremely strong
viewing-angle dependence.
The strong viewing-angle dependence in double detonation
models has previously been discussed \citep[e.g.][]{kromer2010a, gronow2020a}.
The dependence is due to a combination
of line blanketing effects, as well as the distribution
of $^{56}$Ni within the model ejecta.
For example, in Models M10\_05 and M08\_03,
in the southern hemisphere
the $^{56}$Ni is nearer to the 
surface of the ejecta, leading to brighter and earlier
maxima in the bluer bands.
This effect of the distribution of $^{56}$Ni in the model
has previously been discussed by \cite{sim2012a}.
In the northern hemisphere, the $^{56}$Ni is
further from the ejecta surface, which leads to fainter
light curves in these lines of sight.
In addition to this, higher abundances
of heavy elements (such as Ti, Cr and Fe-group elements)
are synthesised in these lines of sights in the outer ejecta,
leading to stronger absorption in the bluer bands,
and in the extreme cases near the pole,
leads to line blanketing of blue wavelengths.
The reverse is true for Model M10\_10,
which was ignited by the edge-lit mechanism.
In this case the $^{56}$Ni is nearer the surface in the 
northern hemisphere (see Figure~\ref{fig:composition}), resulting in these lines of sight showing the brightest light curves.
Despite this reversal, the viewing angle dependence shown by this model
is similar to the level shown by Model M10\_05.

The viewing angle dependence shown by the models
is not as strong in the redder bands, as can
be seen in the V and R bands in Figure~\ref{fig:lightcurvesM10_05}.
We also note that in the bluer bands the viewing 
angle dependence decreases over time.
This can be understood by the ejecta becoming more
optically thin over time, leading to the reduced viewing angle dependence.

\subsubsection{Width-luminosity relation}
\label{sec:viewingangles-Bmax-deltam15B-paramstudy}

\begin{figure}
    \begin{center}
        \includegraphics[width=0.5\textwidth]{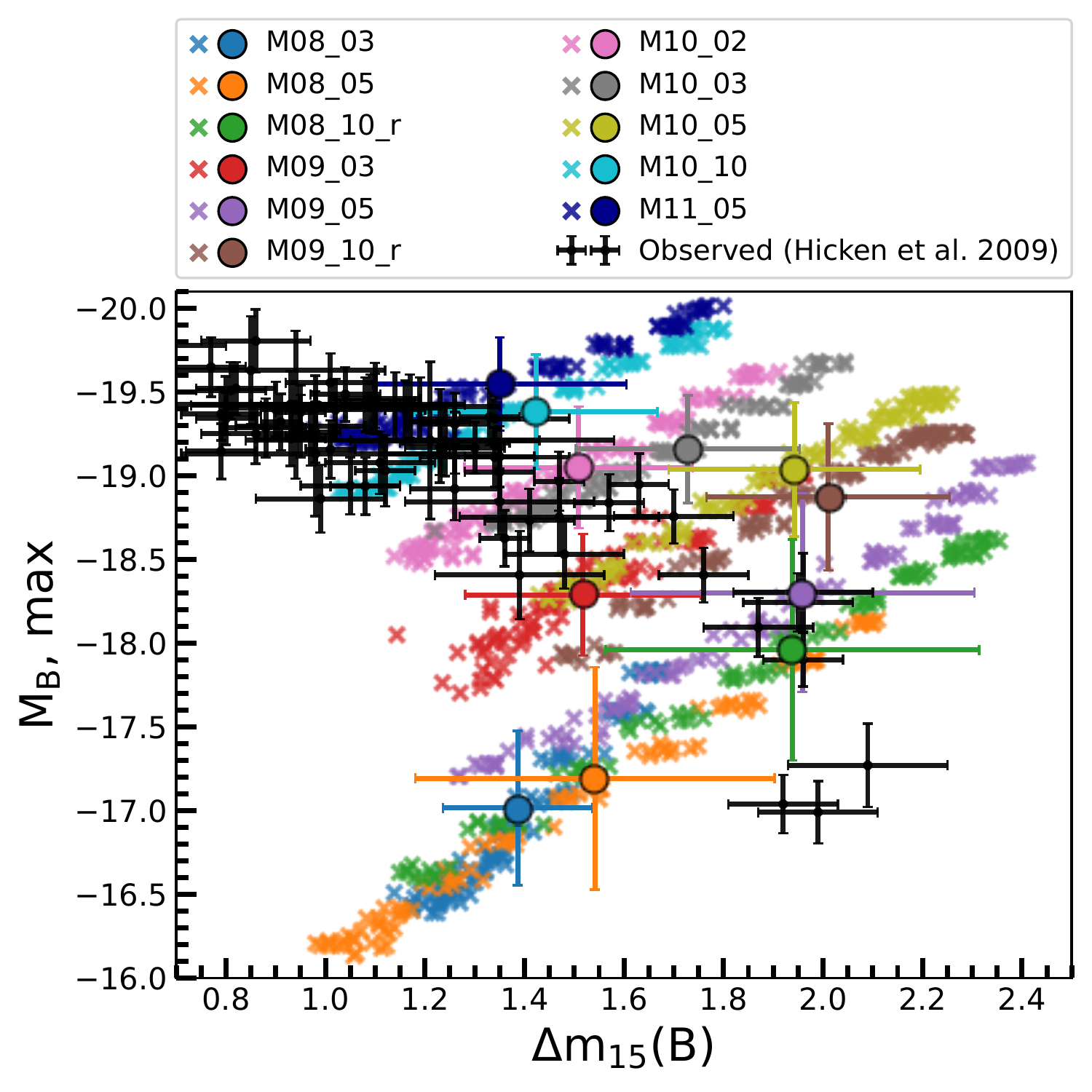}
    \end{center}
    \caption{Peak B band magnitude plotted against $\Delta$m$_{15}$(B) for the angle-averaged
    light curve values (circles) and 100 different viewing angles (x) for each model,
    compared to a sample of observed SNe Ia from \citet{hicken2009b}.
    SNe Ia with a distance modulus $\mu < 33$ have been
    excluded.
    The error bars show the standard deviation of the viewing angle distributions.}
    \label{fig:lightcurvesBbandDeltaM15paramstudy}
\end{figure}

It has previously been shown that sub-\Mch white dwarfs
show reasonable agreement with the observed width-luminosity
relation \citep{sim2010a, blondin2017a, shen2018b, shen2021a} shown by SNe~Ia \citep{phillips1993a}.
This relation can be seen in Figure~\ref{fig:lightcurvesBbandDeltaM15paramstudy},
showing the observed sample of SNe~Ia from \cite{hicken2009b}.
Chandrasekhar mass models for SNe~Ia generally fail to reproduce 
the full width-luminosity relation, such that models
can not explain observations with $\Delta$m$_{15}$(B) $\gtrsim 1.5$
\citep{kasen2009a, sim2013a, blondin2017a}.
Sub-\Mch \ models, however, have been shown to explain
the fainter, faster evolving end of the width-luminosity relation
\citep{sim2010a, blondin2017a, shen2018b, shen2021a}.
The variation in mass readily explains the variation in the 
amount of $^{56}$Ni synthesised, and therefore the variation
in observed peak brightness.
Most previous studies investigating the width-luminosity relation
for sub-\Mch explosion models have considered bare carbon-oxygen
white dwarfs as the explosion models, and therefore did not consider
the mechanism responsible for igniting the detonation.
Here we investigate whether double detonation
models can also explain the variation shown by the width-luminosity relation.
In Figure~\ref{fig:lightcurvesBbandDeltaM15paramstudy}, we plot B band maximum against $\Delta$m$_{15}$(B)
for each model line of sight, as well as the angle-averaged values.
The models indeed show decline rates $\Delta$m$_{15}$(B)$ \gtrsim 1.5$,
and the angle averaged points in Figure~\ref{fig:lightcurvesBbandDeltaM15paramstudy}
generally lie close to the observed SNe~Ia.
However, the extent of the viewing angle dependence is
much greater than that shown by the observational data.
\cite{gronow2021a} similarly found that the bolometric
light curves showed a greater viewing angle dependence than
the bolometric dataset of SNe~Ia of \cite{scalzo2019a}.
Our models, however, do not account for the brighter
end of the observed width-luminosity relation,
showing slower decline rates (see Figure~\ref{fig:lightcurvesBbandDeltaM15paramstudy}).

Despite the observed width-luminosity relation being a characteristic
feature of SNe Ia,
the physical reason for the observed trend is not yet fully understood,
or entirely reproduced by simulations.
The B band width-luminosity relation
is dependent on the diffusion time, but also on the colour evolution of SNe Ia \citep{kasen2007a}.
Simulations of colour evolution are particularly uncertain, especially in the bluer bands,
since transitions at these wavelengths are extremely
sensitive to the micro-physics describing the state of the gas.
Therefore, to simulate the colour evolution, we must have a
good description of e.g. ionisation, excitation and temperature
of the ejecta.
It is likely that the width-luminosity relation of our models
is sensitive to non-LTE effects, and therefore
future full non-LTE simulations may show better agreement
with observations.

In addition to the challenges of simulating colour evolution,
\textsc{artis}, and other codes that do not include full non-LTE treatments,
(see e.g.\ \citealt{shen2018b}) tend to
overestimate the decline rate.
Therefore it is possible that these explosion models would show a slower decline from maximum
if we included a full non-LTE treatment{, e.g. \citet{shen2021a} found their non-LTE models showed a slower B-band 
decline rate than their LTE simulations (a maximum difference of $\Delta$m$_{15}$(B) $= 1$ mag)}.
Full non-LTE simulations are required to quantify this effect {for double detonation simulations}. 
\cite{blondin2017a} use the non-LTE radiative transfer code \textsc{cmfgen},
and find that the sub-\Mch models they consider are able to reproduce
the full width luminosity relation with sub-\Mch explosion models.
Similar results were found by \cite{shen2021a}.
These, however, did not consider the ignition mechanism, and therefore
did not suffer from the apparent problems introduced
by the helium shell detonation, and were also 1D and neglected potential viewing-angle dependencies.
Future work should investigate the effect of full non-LTE simulations
on the width-luminosity relation for double detonation simulations.

The lowest mass models in our study (M08\_03 and M08\_05)
show similar B~band peak brightnesses to faint, 91bg-like objects,
which have peak magnitudes
of~$\sim-17$ mag. They do not, however, account for the decline rates of 91bg-like objects of~$\sim1.9$ (see Figure~\ref{fig:lightcurvesBbandDeltaM15paramstudy}).
Models M08\_03 and M08\_05 do not follow the observed
width luminosity relation.
\cite{gronow2021a} also found that these two models did not
agree with observations in bolometric light.
This is in agreement with \cite{blondin2017a}, 
who found that their low mass (0.88 M$_\odot$) 1D bare white dwarf
detonation model showed an anti-width luminosity relation.
\cite{shen2018b} and \cite{polin2019a} have also noted this,
and following discussion by \cite{shen2014a}, they suggest that this may be due to a
physical minimum white dwarf mass, associated with the central density that can be
ignited via the converging shock mechanism.

\begin{figure}
    \begin{center}
        \includegraphics[width=0.5\textwidth]{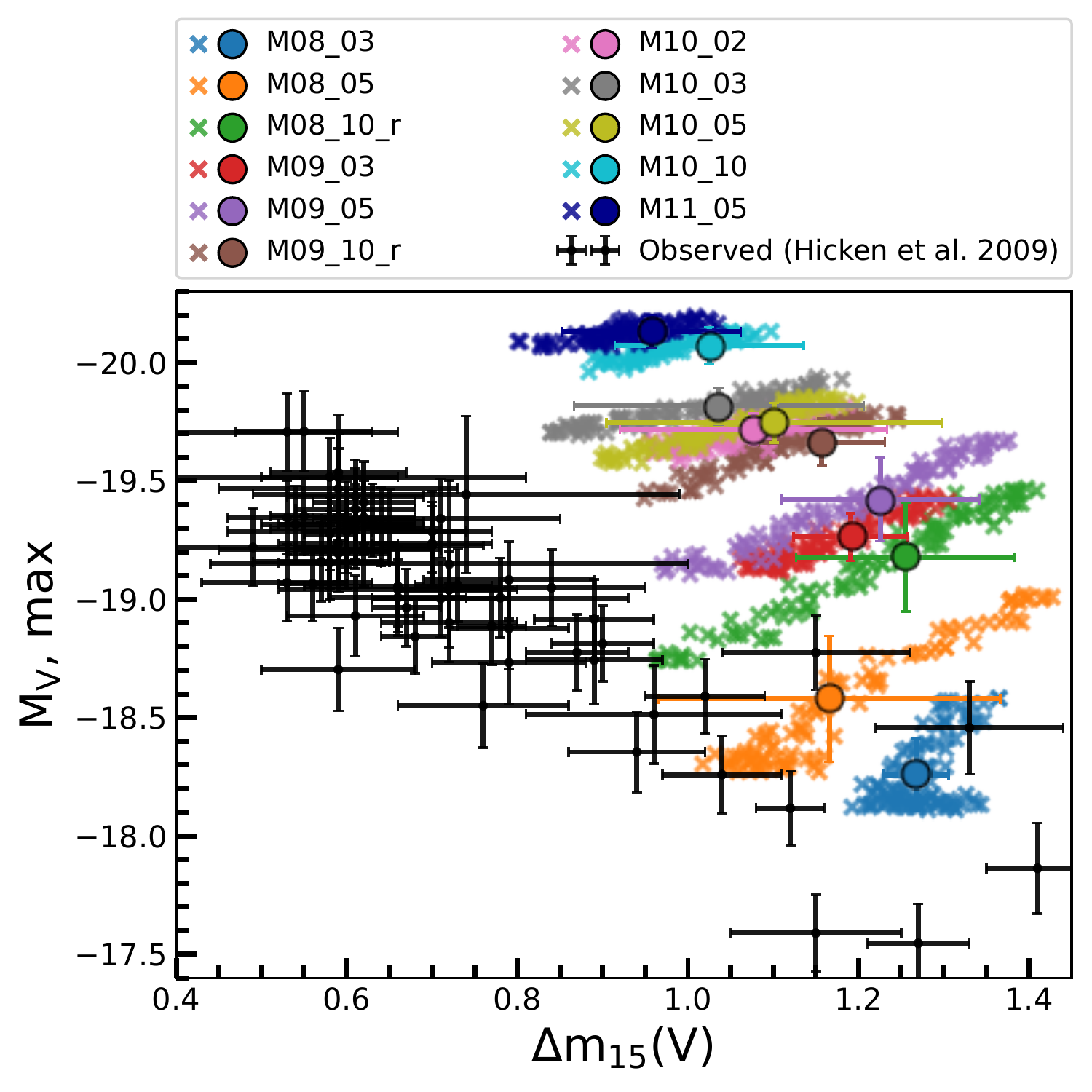}
    \end{center}
    \caption{V-band width luminosity relation. Circles show the angle-averaged
    values while crosses show values in individual lines of sight.
    The error bars on the models show the standard deviation of the viewing angle distribution.
    Also plotted is the sample from \citet{hicken2009b}.
    }
    \label{fig:widthluminosity_Vband}
\end{figure}

Since the B-band light curves are particularly strongly impacted by
absorption from the helium shell detonation ash,
we also show the V-band width luminosity relation
for our models in Figure~\ref{fig:widthluminosity_Vband},
and compare this to observations.
As expected from the light curves, the viewing angle dependence
is not as strong in V-band, particularly for the brighter models.
However, we note
that our V-band light curves appear to be
too bright compared to observations in
Section~\ref{sec:lightcurves-angleave-bandlimited-paramstudy},
and this can be seen in Figure~\ref{fig:widthluminosity_Vband}.
The V-band decline is also too fast compared to observations,
such that we find a systematic offset between models
and observations.

\subsubsection{Rise time}

\begin{figure}
    \begin{center}
        \includegraphics[width=0.5\textwidth]{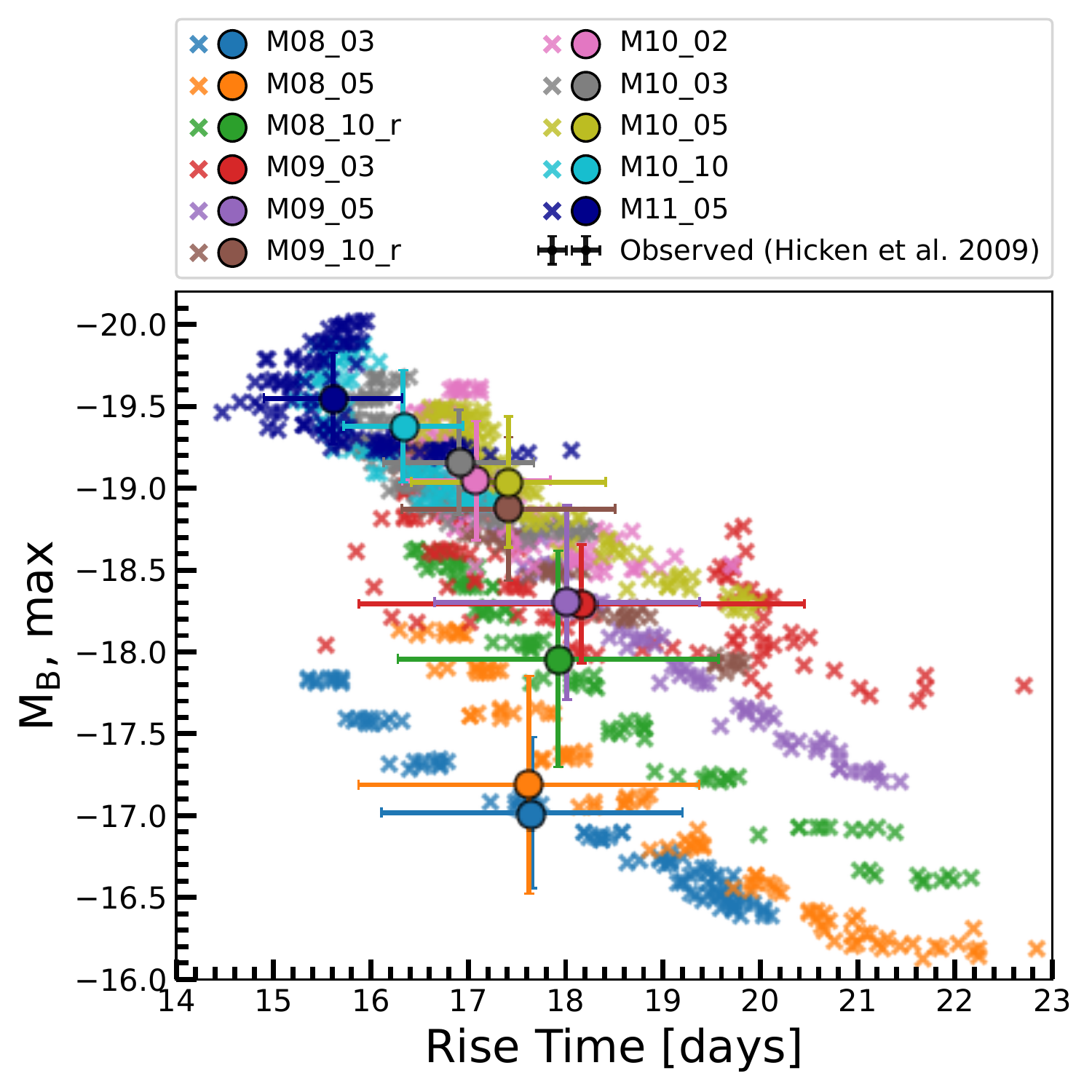}
    \end{center}
    \caption{Time taken to reach B-band maximum light from the time of explosion,
    plotted against peak absolute magnitude in B-band.
    We show the rise time for the angle-averaged light curves (circles)
    as well as in each line line of sight (x).
    The error bars show the standard deviation of the viewing angle distributions.}
    \label{fig:risetime}
\end{figure}

In Figure~\ref{fig:risetime} we show the time taken from
explosion for each model to reach maximum light on average 
and in each line of sight.
Brighter models tend to show a faster rise-time to maximum light,
except for the faintest part of our model range.
The average rise time of our models ranges from 15.7 to 18.1 days.
Apart from Model M11\_05, which showed the fastest rise time, this is
in agreement with rise times found for SNe Ia by \cite{firth2015a}
ranging from 15.98 to 24.7 days.
Particularly in the faster rising lines of sight, Model M11\_05
shows rise times too fast compared to observations.
The variation in rise-time found in each line of sight is primarily driven
by the distribution of $^{56}$Ni in the ejecta, such that
lines of sight where the $^{56}$Ni is nearer the surface 
show a faster rise to maximum.
This can also be seen in Figure~\ref{fig:lightcurvesM10_05}
showing line of sight dependent model light curves.

\subsubsection{B-V colour at B band maximum}
\label{sec:viewingangle_B-V_colour}

\begin{figure}
    \includegraphics[width=0.5\textwidth]{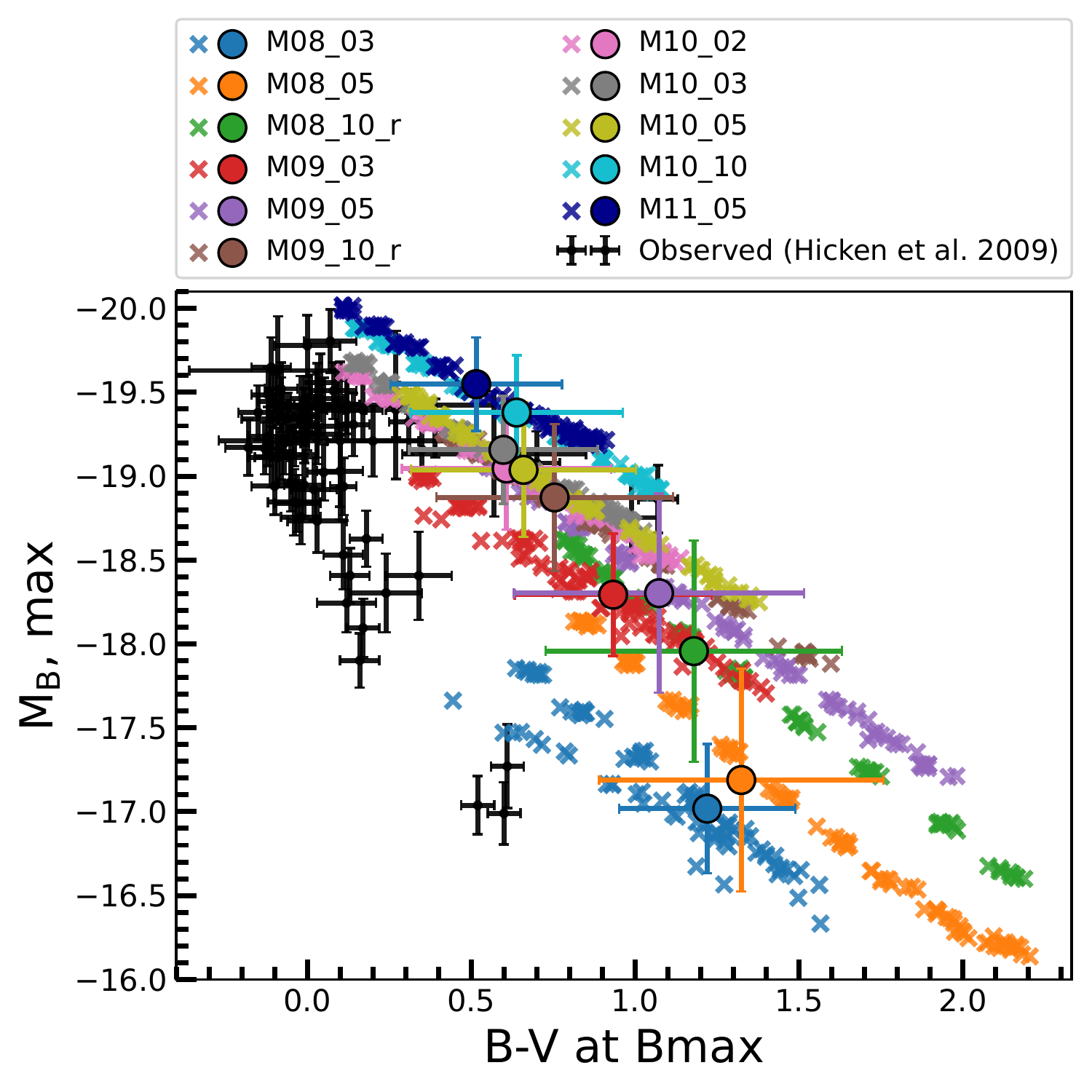}
    \caption{B-V colour at B max for each line of sight, plotted against B band maximum. Also plotted is the observed sample from \citet{hicken2009b}.
    All of the models are too red compared to observations.
    }
    \label{fig:B-VatBmax}
\end{figure}

In Figure~\ref{fig:B-VatBmax} we show the B-V colour at the time of B band maximum
for each of the 100 different viewing-angles in the 3D models.
Again this highlights the extent of the viewing angle dependence in
our set of double detonation simulations.
We also show the colour of a sample of normal SNe Ia from \cite{hicken2009b}.
It is clear from this that all of our models are too red compared to
normal SNe Ia.
In the brighter, more massive models, the brightest lines of sight lie within the range of observations,
however overall, even for these models, the colours are too red.
This includes our model with the lowest mass helium shell,
Model M10\_02.
As previously discussed, light curve colour is challenging 
to simulate, given the dependence of wavelength on the conditions
in the ejecta, such as temperature, ionisation and excitation state, as well as dependence on the atomic data included in the calculation.
In particular, the bluer bands are sensitive to this, given the number
of transitions at these wavelengths.
The large abundances of heavy elements produced in the outer ejecta layers
in the helium detonation cause strong absorption of bluer wavelengths in
our simulations, leading to the extremely red colours.
However, this absorption may be exaggerated by the approximations 
made in the treatment of ionisation in our radiative transfer calculations.
We note that colours redder than observations have previously been found
by \textsc{artis} for other classes of explosion models,
including delayed-detonation models \citep{sim2013a} and
pure detonations \citep{sim2010a}, however, these colours were only slightly
redder than observations (B-V at maximum $\approx 0.5$ for the reddest models),
and not the extremely red colours found here.
Full non-LTE simulations are required to fully investigate the B-V colour
of these double detonation models, and determine whether this is a problem with
the explosion scenario, or due to shortcomings in the radiative transfer simulations.

\subsection{Spectra}
\label{sec:param-study-spectra}

In this section we present the model spectra.
We first discuss the angle-averaged spectra in Section~\ref{sec:spectra-angleave-paramstudy},
and then make comparisons
of the line of sight spectra to observations
in Sections~\ref{sec:viewinganglespectra} and \ref{sec:paramstudy-comparetoSN2018byg}.

\subsubsection{Angle-averaged spectra}
\label{sec:spectra-angleave-paramstudy}

\begin{figure}
    \begin{center}
        \includegraphics[width=0.5\textwidth]{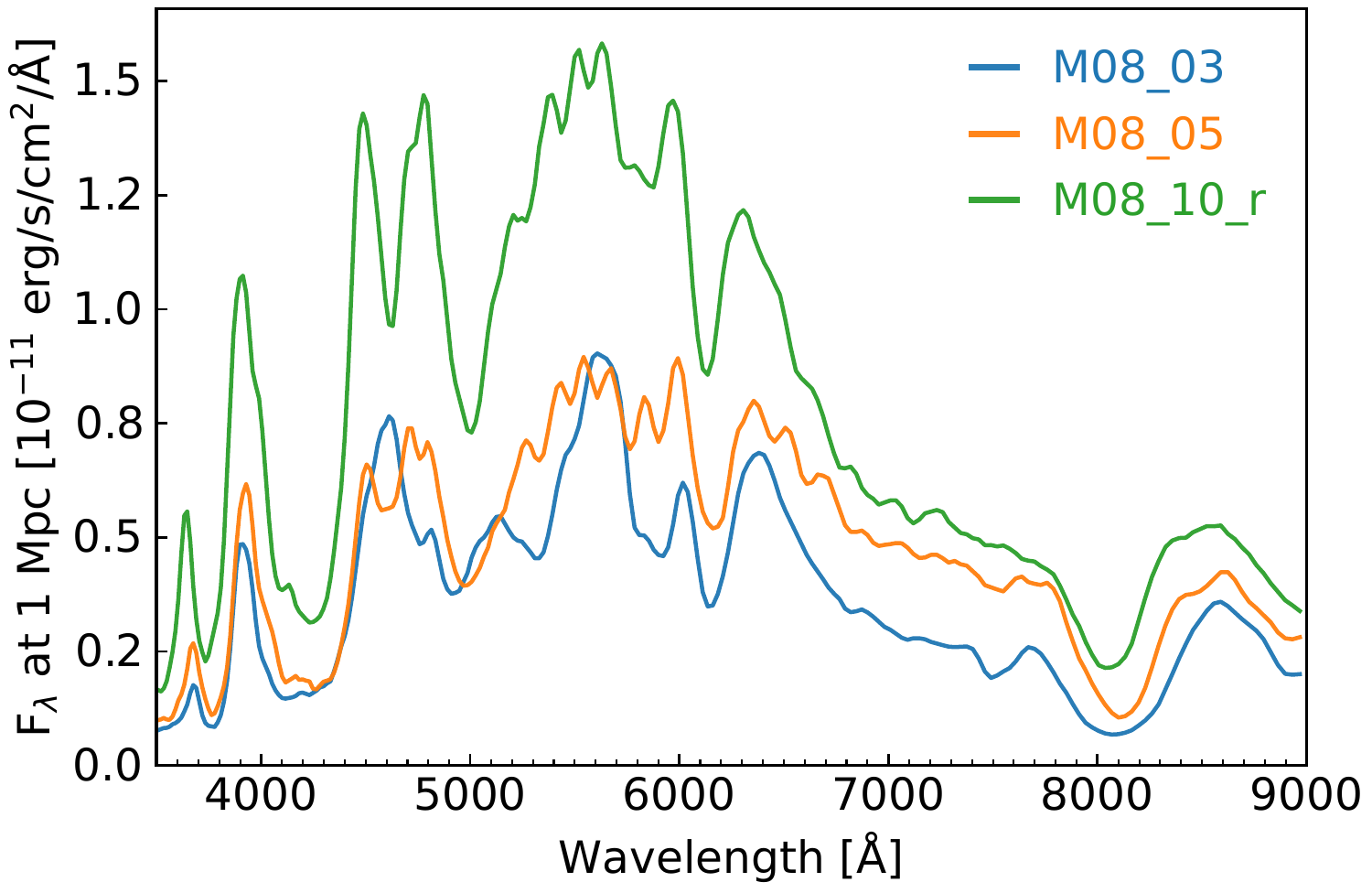}
        \includegraphics[width=0.5\textwidth]{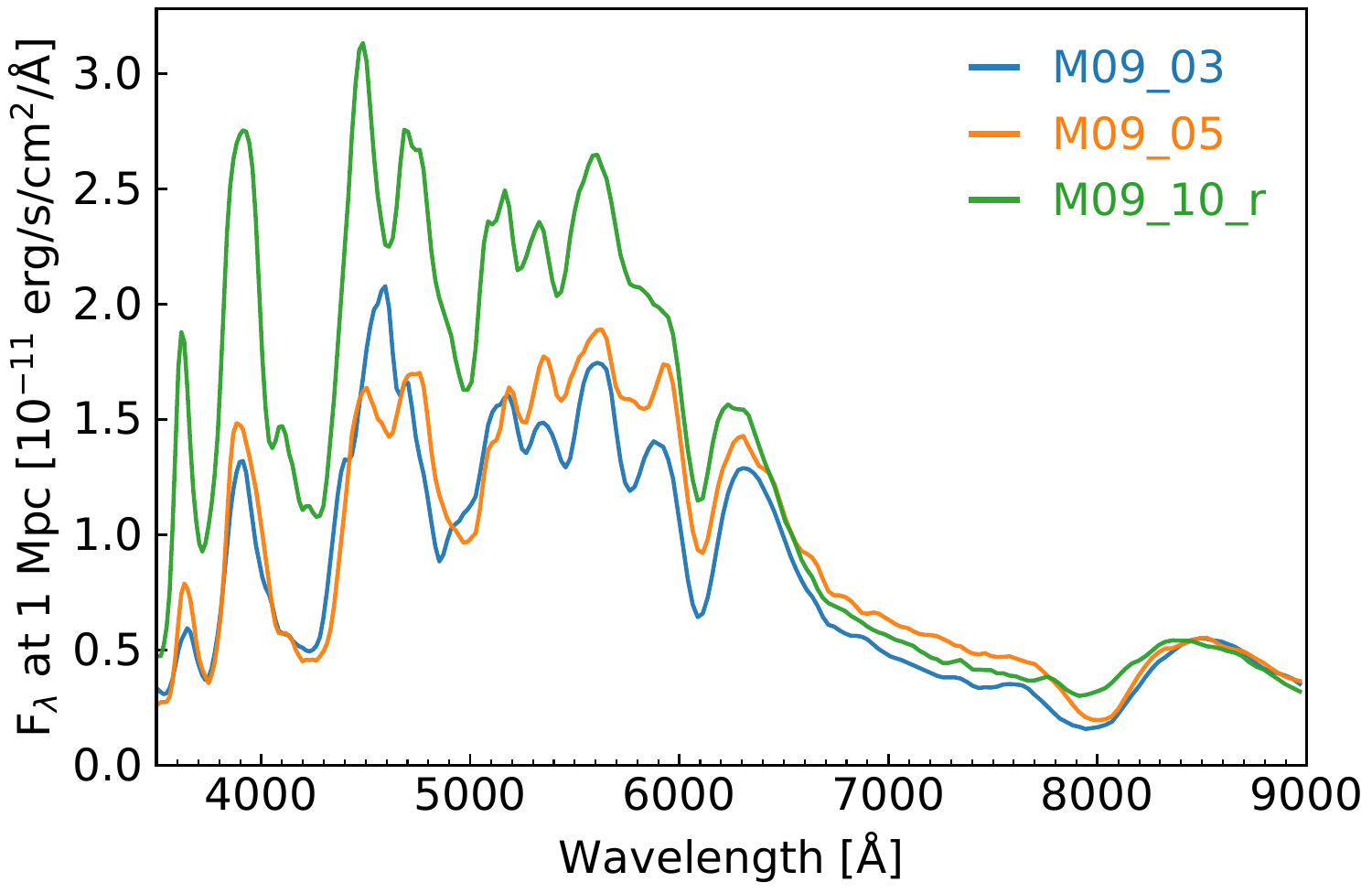}
        \includegraphics[width=0.5\textwidth]{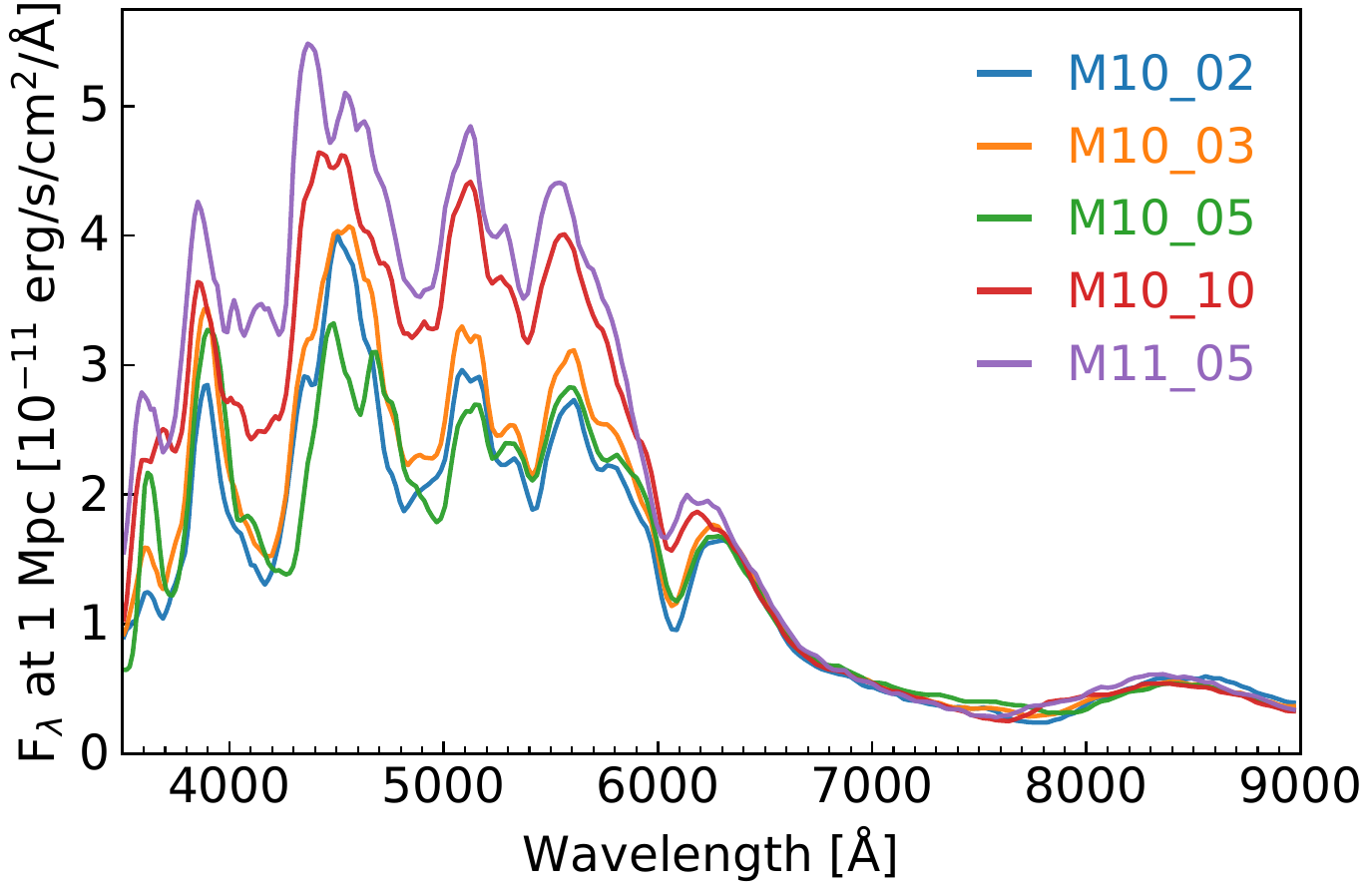}
    \end{center}
    \caption{Angle-averaged model spectra at 2 weeks after explosion.
    }
    \label{fig:spectra-angleave-paramstudy}
\end{figure}
\begin{figure}
    \begin{center}
        \includegraphics[width=0.5\textwidth]{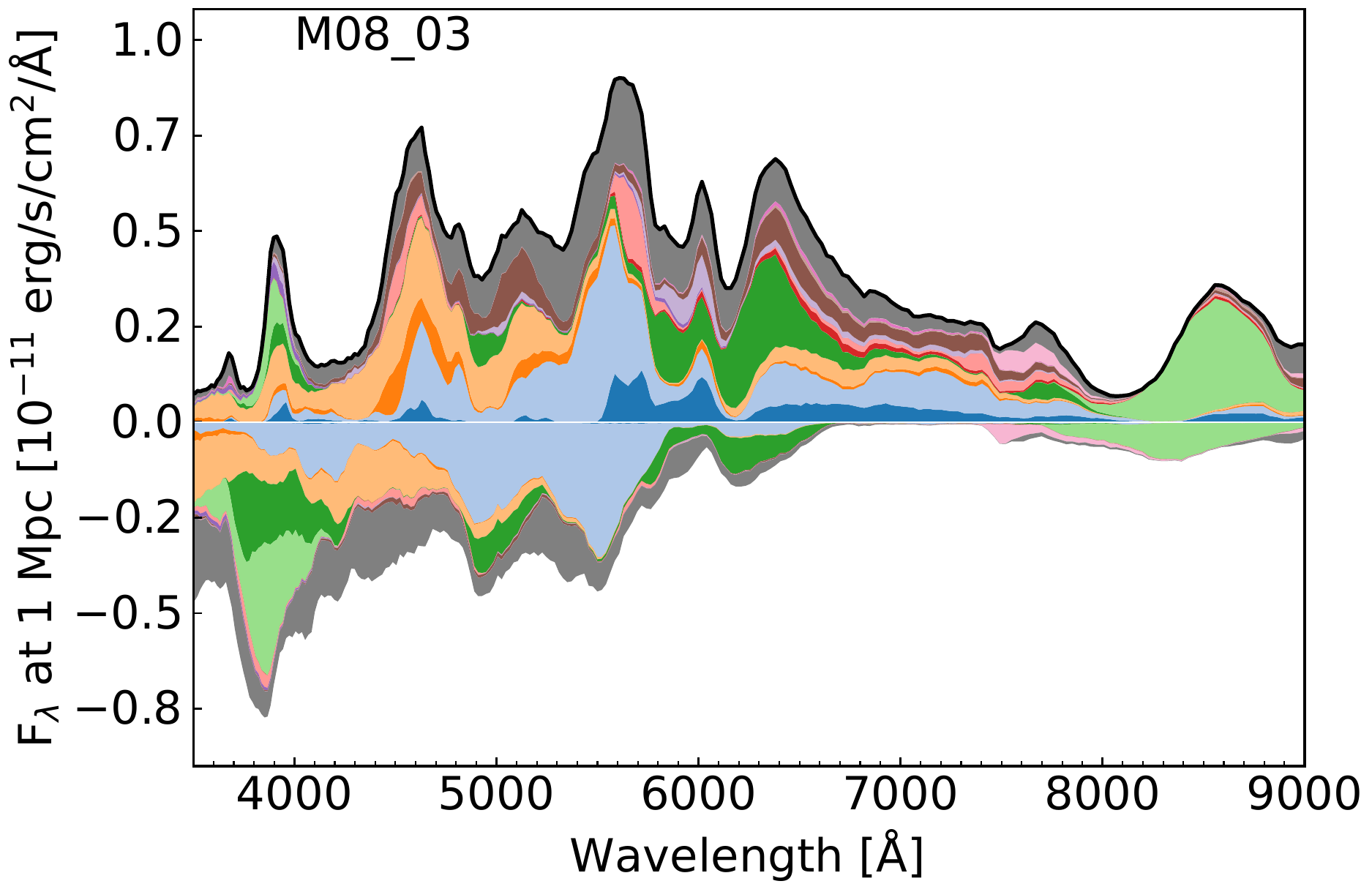}
        \includegraphics[width=0.5\textwidth]{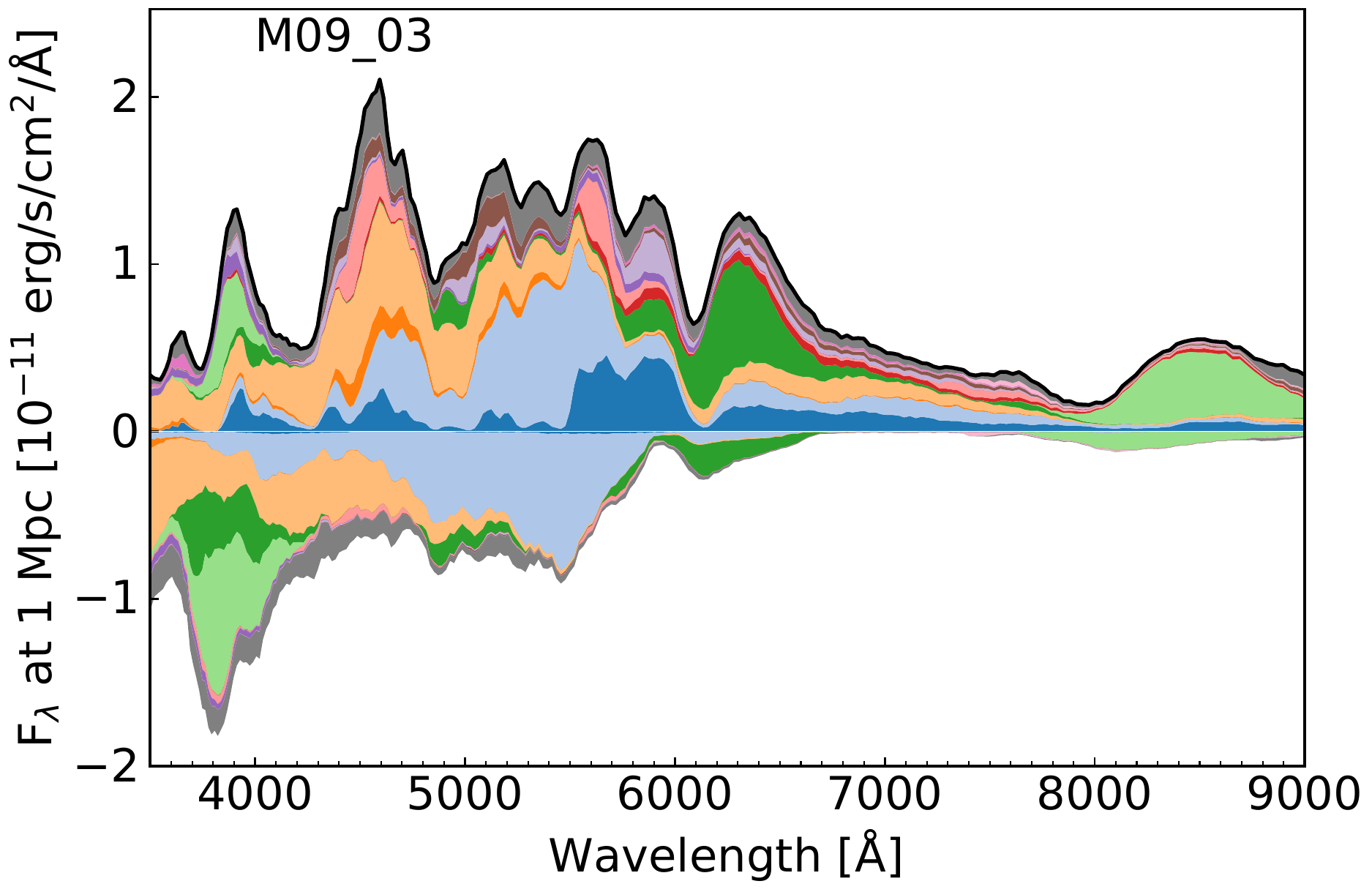}
        \includegraphics[width=0.5\textwidth]{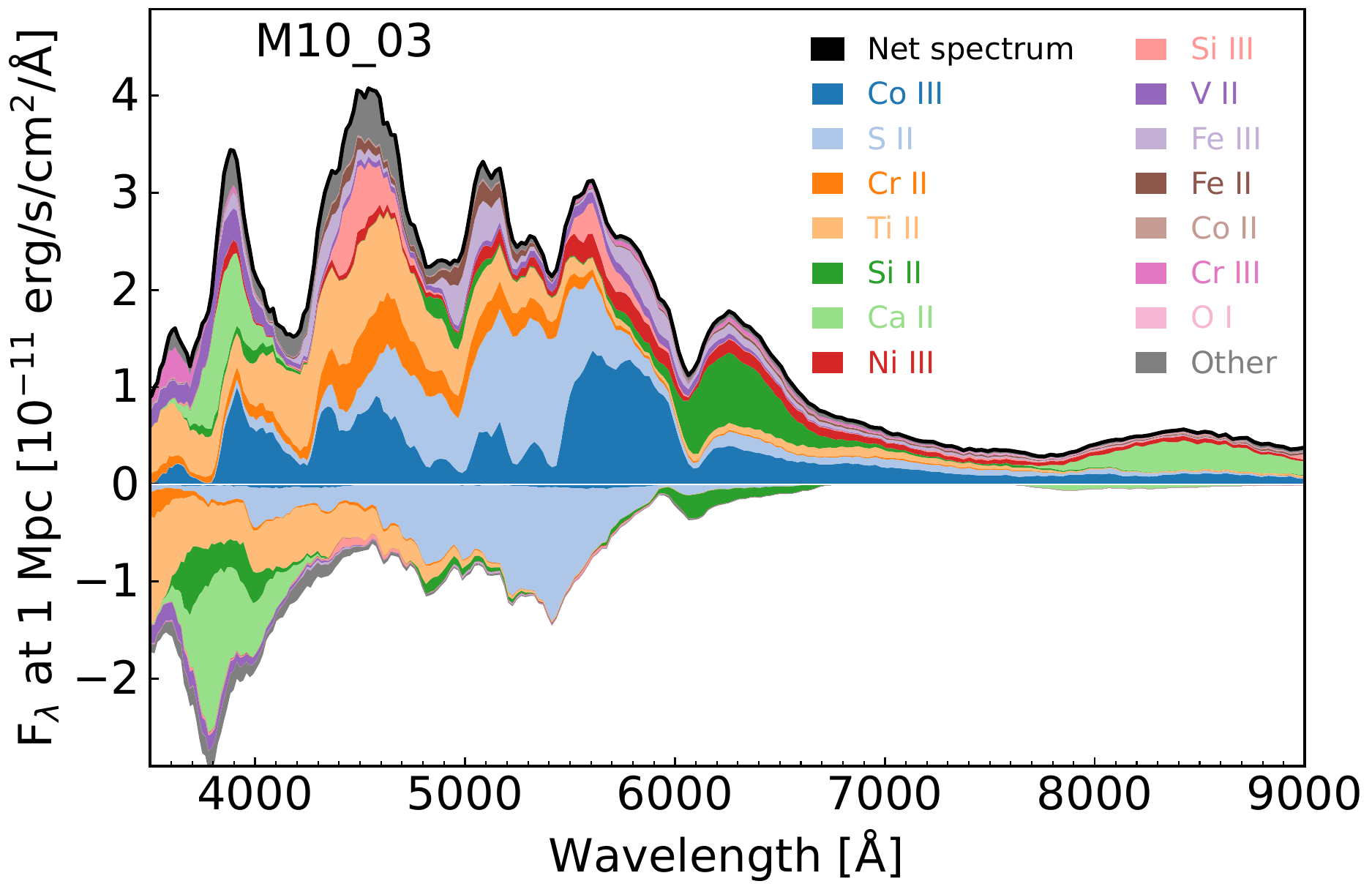}
    \end{center}
    \caption{Spectra at 2 weeks after explosion for each of the models
    with helium shells of $\sim$ 0.3 M$_\odot$,
    where the contributions by individual ions to the spectra are indicated.
    Beneath the axis we also indicate the ions responsible for absorption.
    These can be directly compared to Figure~\ref{fig:spectra-angleave-paramstudy}.
    }
    \label{fig:emissionabsorption}
\end{figure}

We show the model angle-averaged spectra at two weeks after explosion in
Figure~\ref{fig:spectra-angleave-paramstudy}, and compare models
with similar total masses.
All of the spectra show characteristic IME spectral features,
including \ion{Si}{II} $\lambda 6355$, \ion{S}{II} $\lambda \lambda 5454, 5640$ and the \ion{Ca}{II} triplet at $\lambda \lambda$8498, 8542, 8662.
However, despite relatively low masses of Ti in our models,
all of the model spectra show a clear absorption
feature at $\sim$ 4000--4500 \AA, dominated by \ion{Ti}{II} absorption.
Such a feature is observed in sub-luminous, 91bg-like SNe Ia,
due to the lower temperatures, and therefore lower ionisation state of the
ejecta \citep{mazzali1997a}, but is not observed in normal SNe Ia.
We note however that such a feature has been observed in the normal
brightness, peculiar SN~2016jhr \citep{Jiang2017a}, and shows good
agreement with the spectral features predicted for the double detonation
Model M2a of \cite{gronow2020a}, which is similar to Model M10\_05.
The prediction of a \ion{Ti}{II} absorption feature 
for the double detonation is in agreement with the previous
findings of \cite{kromer2010a}, \cite{townsley2019a} and \cite{gronow2020a}.
As model mass decreases, and therefore as model brightness decreases,
we find stronger \ion{Ti}{II} absorption and redder colours.
Since the lower mass models tend to have higher abundances of Ti,
and on average lower temperatures and ionisation states, this is to be expected.

Even our lowest helium shell mass model (M10\_02) shows a clear \ion{Ti}{II} absorption feature, which is similar
in strength to the \ion{Ti}{II} feature predicted for models
M10\_03 and M10\_05 (see Figure~\ref{fig:spectra-angleave-paramstudy}).
\cite{townsley2019a} have suggested that thin helium shells of
similar mass to that in Model M10\_02 can produce spectra
of normal SNe Ia, however, we find that even our lowest helium shell mass
model still predicts spectra that would be classified as a peculiar SN Ia.
In Figure~\ref{fig:emissionabsorption} we indicate the
key ions contributing to the emission spectrum for each of Models M10\_03,
M09\_03 and M08\_03 at 2 weeks after explosion,
and also indicate the species responsible for absorption
(shown beneath the axis).

\subsubsection{Viewing-angle dependent spectra}
\label{sec:viewinganglespectra}

As demonstrated in Section~\ref{sec:viewingangle-lightcurves},
the models show strong viewing-angle dependencies,
particularly at bluer wavelengths, where the products of the 
helium shell detonation most strongly affect the spectrum.
We now discuss the viewing-angle dependence shown by the spectra.

\begin{figure}
    \begin{center}
        \includegraphics[width=0.5\textwidth]{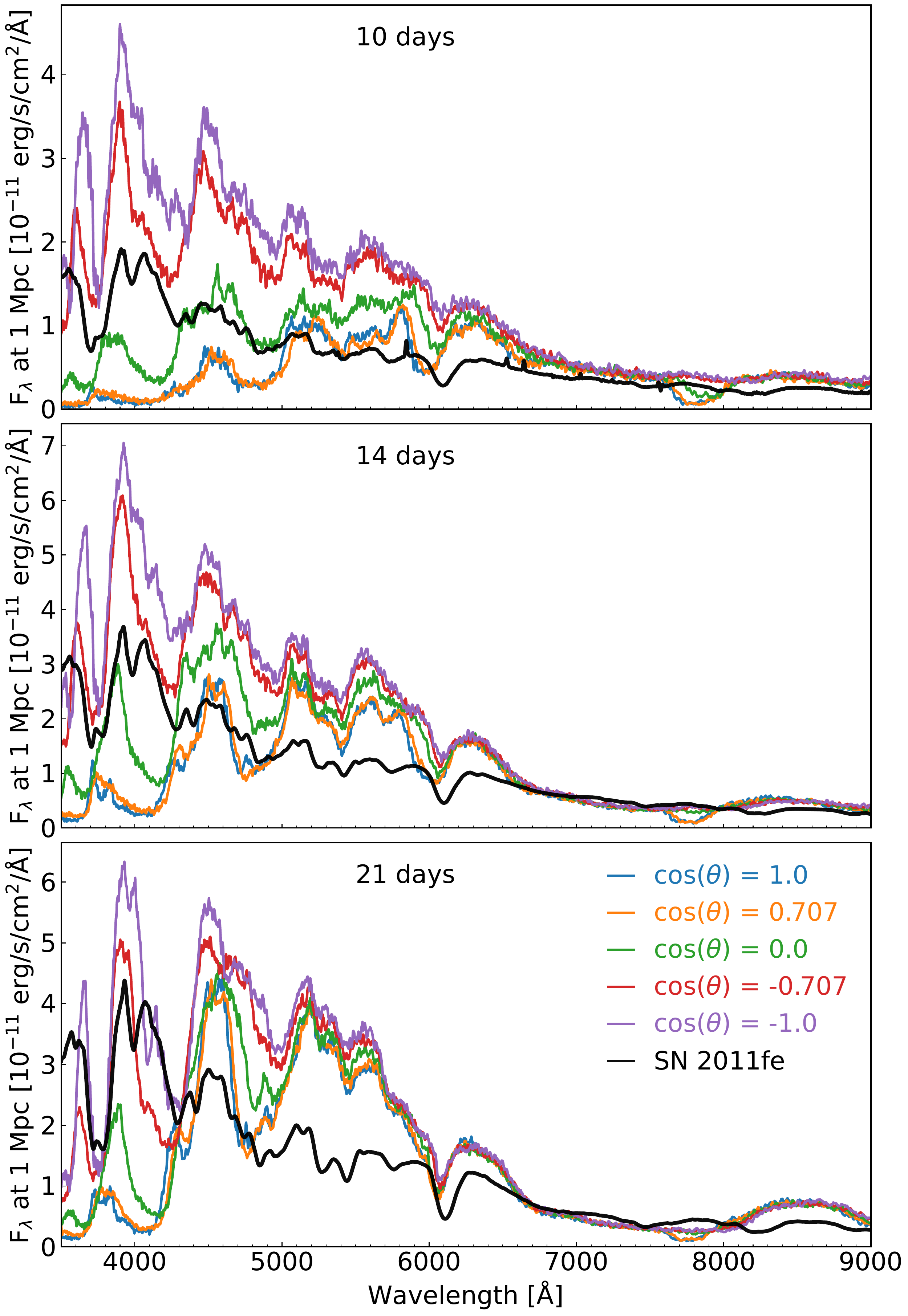}
    \end{center}

    \caption{Spectra in specific lines of sight of
    Model M10\_03. These were calculated using `virtual packets'.
    We compare the model spectra to SN~2011fe at similar
    epochs.}
    \label{fig:viewinganglespectra_M10_03}
\end{figure}

Figure~\ref{fig:viewinganglespectra_M10_03} shows spectra
in specific lines of sight
for Model M10\_03.
As expected, given the asymmetries shown by the light curves,
the spectra at 2 weeks after explosion show a strong viewing
angle dependence.
Lines of sight viewing towards the northern pole,
where the helium detonation was ignited at cos($\theta$) = 1,
show extreme line blanketing at blue wavelengths,
as has previously been discussed by \cite{kromer2010a}
and \cite{gronow2020a}.
In these lines of sight, high abundances of heavy elements were
produced in the helium shell detonation, such as
Cr, Ti and Fe-peak elements.
These elements are extremely effective at line blanketing.
In addition to the high abundances of heavy elements,
the $^{56}$Ni was produced farthest 
from the surface in these lines of sight.
Therefore it is not unexpected that the spectra in these
lines of sight are fainter, however, consequently, these
are also the coolest and least ionised,
which has the effect of increasing the level of line blanketing.
We note that redder wavelengths do not show a strong 
viewing angle dependence, except for the \ion{Ca}{II} triplet,
which we find to be stronger and at higher velocities
in lines of sight near cos($\theta$) = 1.

In Figure~\ref{fig:viewinganglespectra_M10_03}, we also plot spectra
of the normal SN~Ia, SN~2011fe, at similar epochs.
At each epoch our model spectra produce characteristic SNe Ia features,
such as \ion{Si}{II} $\lambda$6355 and \ion{S}{II} $\lambda \lambda 5454, 5640$.
However, in all lines of sight the spectra are much redder than the
spectra of SN~2011fe, which represents a typical `normal' Type Ia SN.
Additionally, the model \ion{Si}{II} $\lambda$6355 features are produced at higher velocities than the observations.
We note that the spectra of Model M10\_02 are similar
to model M10\_03.

\subsubsection{Line of sight spectra compared to SN~2018byg}
\label{sec:paramstudy-comparetoSN2018byg}

SN 2018byg was suggested by \cite{de2019a} to be the result of a 
double detonation.
The bluer regions of the spectra of
SN~2018byg showed unusually strong line blanketing, with broad \ion{Ti}{ii} and
Fe-group element absorption features,
and near peak a deep, high
velocity ($\approx$ 25 000 km s$^{-1}$) \ion{Ca}{ii} triplet absorption
feature was observed.
The light curves of SN~2018byg are subluminous and similar to
SN~1991bg-like SNe~Ia, except for a rapid rise in r-band magnitude
within the first week from explosion.
\cite{gronow2020a} compared the level of absorption in the faintest lines of sight
in their reference model, M2a, to the absorption shown by SN~2018byg, and found that this was a good match to the
level of line blanketing observed.
However, Model M2a showed light curves brighter than those observed for SN~2018byg.
We now make this comparison with a model showing peak brightnesses similar to SN~2018byg,
which had a peak absolute magnitude of M$_{\mathrm{R, max}} \approx -18.2$ mag~\citep{de2019a}.
This is similar to Model M08\_03 which showed M$_{\mathrm{R, max}} = -18.4$ mag.

\begin{figure}
    \begin{center}
        \includegraphics[width=0.5\textwidth]{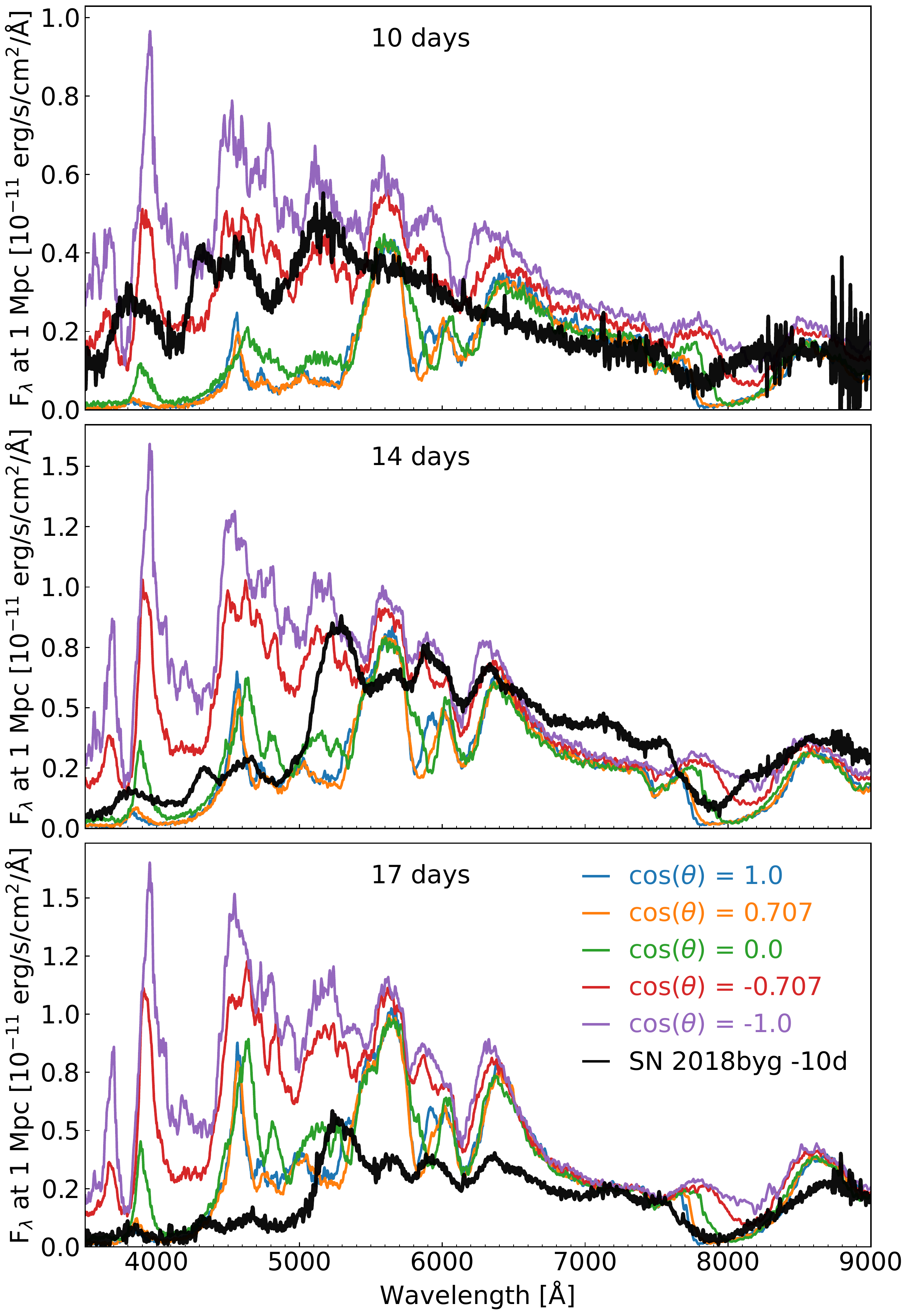}
    \end{center}

    \caption{Spectra of Model M08\_03 in specific lines of sight at 10,
    14 and 17 days after explosion. Also shown are the spectra of SN~2018byg at similar epochs.}
    \label{fig:spectra_viewingangles_M08_03}
\end{figure}

In Figure~\ref{fig:spectra_viewingangles_M08_03} we show
the spectra of M08\_03 compared to spectra of SN~2018byg
at similar epochs.
At 10 days after explosion, the model line of sight at 
cos($\theta$)$= -0.707$ is similar to SN~2018byg,
however, by 14 days after explosion, SN~2018byg
shows stronger line blanketing and more closely resembles
the lines of sight at cos($\theta$)$= 1$ and cos($\theta$)$= 0.707$.
By 17 days after explosion the model no longer shows
strong enough line blanketing
to explain that in SN~2018byg.

\begin{figure}
    \begin{center}
        \includegraphics[width=0.5\textwidth]{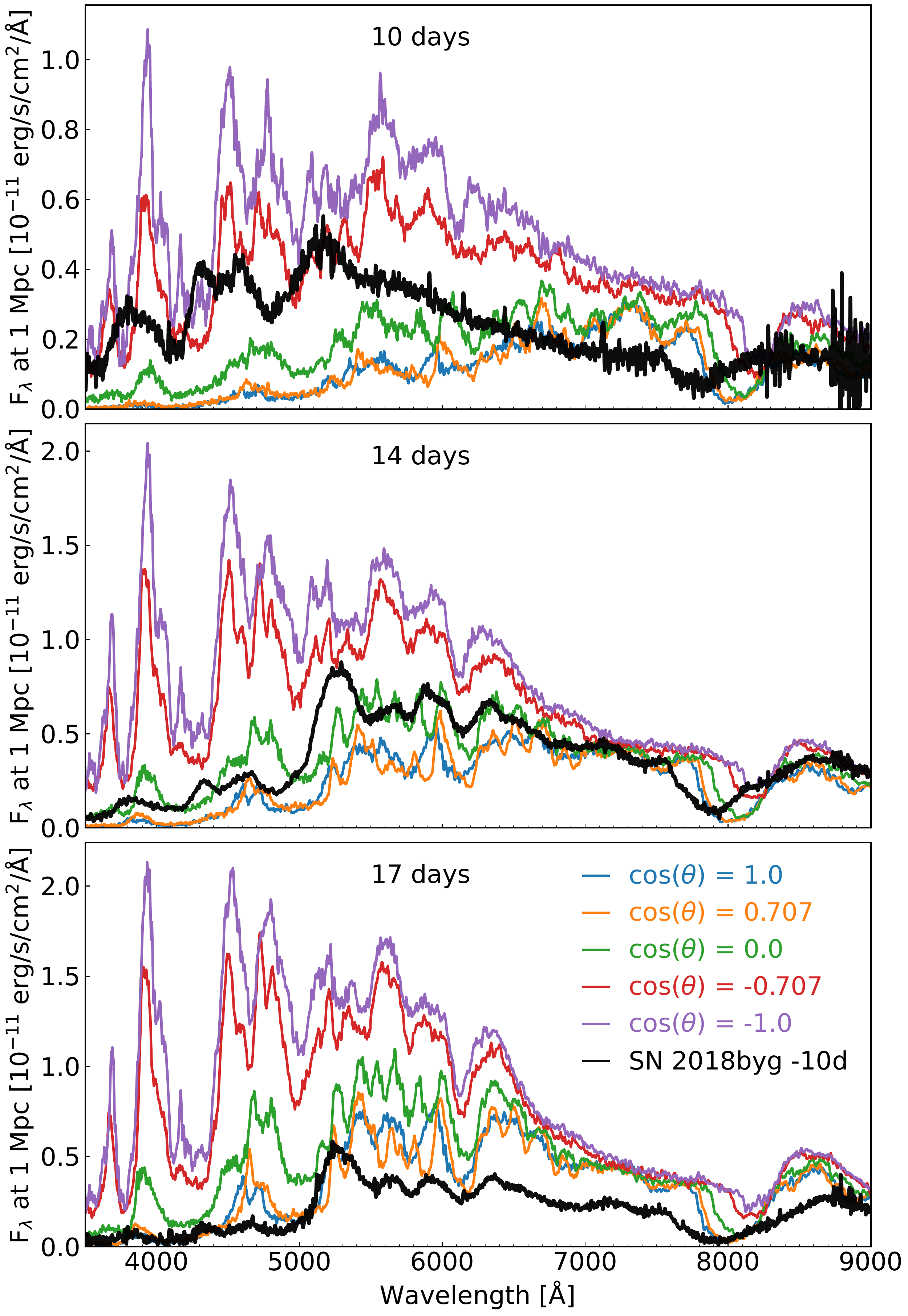}
    \end{center}

    \caption{Same as Figure~\ref{fig:spectra_viewingangles_M08_03}, but compared to the spectra of Model M08\_05.}
    \label{fig:spectra-M08_05-viewingangles}
\end{figure}

Since the model with the lower shell mass does not account
for the line blanketing in SN~2018byg, we
compare SN~2018byg to Model M08\_05 in
Figure~\ref{fig:spectra-M08_05-viewingangles}.
The angle-averaged peak R band magnitude of Model M08\_05 is M$_{\mathrm{R, max}} = -18.8$, although the
faintest lines of sight have M$_{\mathrm{R, max}} = -18.6$.
The R band light curves in our models do not show as strong an angle dependence as shown by the bluer
bands (see Figure~\ref{fig:lightcurvesM10_05}).
We find that the more massive He shell of Model M08\_05 is able to account for the line blanketing observed
in SN~2018byg.
At redder wavelengths, however, Model M08\_05 is brighter than SN~2018byg.
Therefore it is possible that a lower mass core might be able to better account for SN~2018byg.
We also compared the line of sight spectra for Model M08\_10 to SN~2018byg, and
found that the spectrum in the faintest line of sight is able to account for
the line blanketing in SN~2018byg, however, the model spectrum is too bright at redder wavelengths, and does not account
for the broad \ion{Ca}{II} absorption feature.

This suggests that the mass of the He shell of SN~2018byg could have been similar to Model M08\_05.
This is a significantly lower shell mass than suggested by \cite{de2019a} for SN~2018byg:
\cite{de2019a} suggest that a He shell mass of $\approx 0.15$~M$_\odot$ is required to explain the
observed properties of SN~2018byg, whereas the shell mass of
Model M08\_05 was 0.053 M$_\odot$ (pre-relaxation; see \citealt{gronow2021a} for the definition of the He shell mass after mixing was allowed to take place).

\section{Discussion and conclusions}
\label{sec:paramstudy-summary}

We have carried out radiative transfer simulations 
for a series of simulations by \cite{gronow2021a}
of the double detonation explosion scenario,
where the initial core and helium shell masses were varied.

The variation in model mass produced a range of masses of
$^{56}$Ni, and therefore the simulated light curves
for these models showed a range of peak brightnesses.
The brightnesses shown by the light curves were able to
account for normal SNe Ia, as well as fainter 91bg-like SNe.
However, in agreement with previous works, the colours are much redder than observed for normal
SNe Ia for all of our models, due to strong absorption features
by heavy elements produced in the outer ejecta 
during the helium shell detonation.
\cite{townsley2019a} and \cite{shen2021b} found that minimal
helium shell models are able to produce normal SNe Ia,
however, our minimal helium shell model, M10\_02,
still shows colours too red.
This shows that even low masses of He detonation ash can
be significant enough to affect the model colours.

The double detonation explosion models
showed strong asymmetries, which seem to be in conflict
with the tight relationship shown by the Phillips relation.
All three secondary detonation mechanisms found in this study
(converging shock, scissors and edge-lit mechanisms) produced 
similarly asymmetric ejecta.
The bluer band light curves showed an extremely large viewing
angle dependence.
These wavelengths showed strong absorption and 
line blanketing in some lines of sight, due to the helium
shell burning products.
The redder bands, however, did not show such strong asymmetries.
We also found that over time the light curves became less
angle dependent as the ejecta became more optically thin.

In agreement with previous studies of sub-\Mch explosions
\citep{sim2010a, blondin2017a, shen2018b, shen2021a},
the models show variation in decline rate with peak brightness,
which are able to reproduce the overall trend of the faint
end of the width luminosity relation.
However, they do not reproduce the bright end
of the width luminosity relation.
It is possible that future, full non-LTE simulations may change
the predicted decline rates, since the flux in the B band is
sensitive to ionisation state, and it is likely that our approximate
non-LTE treatment underestimates the ionisation at later times.
Simulations by \cite{shen2021a} of sub-\Mch explosion models indicate that the decline rate is affected by considering non-LTE 
{(a slower decline rate of up to $\Delta$m$_{15}$(B) $= 1$ mag for their non-LTE simulation compared to LTE)}.

We also find that the viewing angle dependence in the width-luminosity
shown by the models is stronger than is observed.
The asymmetry of the models is predominately in the helium shell
detonation ash.
If the radiative transfer effects from the helium shell were reduced,
e.g. by considering an even thinner He shell than Model M10\_02,
or through ionisation effects, it is possible that the models
would not show such strong asymmetries.

All of our model spectra show signatures of the helium shell
detonation products, particularly \ion{Ti}{II}, that are not observed in normal SNe Ia.
This includes the minimal He shell mass model, M10\_02.
Similarly to \cite{kromer2010a} and \cite{gronow2020a},
the spectra show a strong viewing angle dependence.
Lines of sight opposite to the ignition point of the helium 
detonation show strong absorption features, 
particularly at blue wavelengths, due to the helium shell
detonation products, which have the highest abundance in these lines of sight.
Additionally, the $^{56}$Ni produced in the CO core detonation
is furthest from the surface of the ejecta in these lines of
sight, and as a result these are cooler than lines of sight where
the $^{56}$Ni from the core detonation is closer to the surface.
Therefore the products of the helium shell detonation are less
ionised and cause stronger absorption features.
It is possible that such lines of sight could explain peculiar
SNe Ia, such as SN 2018byg, which showed strong line blanketing in its
spectra. 
The level of line blanketing observed can be explained by our models
with similar peak brightness.

For double detonations to be able to explain normal SNe Ia,
the effects of the helium shell detonation need to be minimised.
Pure detonation models (where the CO core is detonated without a helium shell)
show good agreement with observations 
(e.g. \citealt{sim2010a, blondin2017a, shen2018b, shen2021a}),
and double detonation explosion simulations with the helium detonation
ash removed (e.g. \citealt{kromer2010a, pakmor2022a}) show similarly 
good agreement with observations, and would no longer be classified 
as peculiar SNe Ia.
Although our lowest mass helium shell model does not reproduce
normal SNe Ia, \cite{townsley2019a} and \cite{shen2021b} have shown that
their low mass helium shell models are able to better reproduce 
normal SNe Ia.
The $1.0$M$_\odot$, $\rho _5 = 3$ model by \cite{boos2021a}, considered
by \cite{shen2021b}, and the model by \cite{townsley2019a}
are similar to our Model M10\_02.
In Table~\ref{tab:Ti-Cr-masses} we compare the masses of $^{44}$Ti and $^{48}$Cr
produced in the helium shell detonations for these models.
Our Model M10\_02 has the highest abundances of these species produced in the shell
detonation, which is likely the primary reason for the differences in colour
found in our simulations.
{The \citet{boos2021a} $1.0$M$_\odot$, $\rho _5 = 3$ model uses
the same model as \citet{townsley2019a} (initial parameters, enrichment, and nuclear reaction network). 
However, for modelling the detonations \citet{boos2021a} used a burning
limiter while \citet{townsley2019a} did not.
This may point to the treatment of the He detonation in the explosion simulations as the reason for the 
differences in abundances.
To investigate this, the He detonation should be spatially resolved, which is within reach for future simulations but beyond the scope of this work.}

\begin{table}
\caption{Masses of $^{44}$Ti and $^{48}$Cr produced in the helium shell detonation
for our Model M10\_02, the 1.0 + 0.02 M$_\odot$ \ model by \citet{townsley2019a}
and the $1.0$ M$_\odot$, $\rho_5 = 3$ from \citet{boos2021a} and \citet{shen2021b}.
For similar initial CO core and He shell masses our model produces the highest
abundances of $^{44}$Ti and $^{48}$Cr in the shell detonation.
These species are predominantly responsible for the red colours found in our radiative
transfer simulations.}
\label{tab:Ti-Cr-masses}
\centering
\begin{tabular}{ccc}
\hline
Model & Shell $^{44}$Ti                                & Shell $^{48}$Cr            \\
      &     [M$_\odot$]                                  &       [M$_\odot$]                     \\ \hline
M10\_02   & $5.7 \times 10^{-4}$               & $2.3 \times 10^{-4}$     \\
1.0 + 0.02 M$_\odot$  & $1.7\times10^{-5}$                 & $2.1\times10^{-6}$       \\
$1.0$ M$_\odot$, $\rho _5 = 3$ & $2.5\times10^{-4}$                 & $1.1\times10^{-5}$       \\ \hline 
\end{tabular}
\end{table}

Although our models do not reproduce normal SNe Ia, we do find that double detonations
remain promising candidates for peculiar SNe Ia showing
strong \ion{Ti}{II} absorption or line blanketing.
Full non-LTE simulations will be important for determining whether the apparent discrepancies found in this
work are due to the explosion modelling, or due to
shortcomings in the approximations made in the radiative transfer simulations.

\section*{Acknowledgements}

CEC acknowledges support by the European Research Council (ERC) under the European Union’s Horizon 2020 research and innovation program under grant agreement
No. 759253.
The work of SG and FKR was supported by the Deutsche Forschungsgemeinschaft (DFG, German Research Foundation) -- Project-ID 138713538 -- SFB 881 (``The Milky Way System'', Subproject A10), by the ChETEC COST Action (CA16117), by the National Science Foundation under Grant No.\ OISE-1927130 (IReNA), and by the Klaus Tschira Foundation. 
The work of SAS was supported by the Science and Technology Facilities Council [grant numbers ST/P000312/1, ST/T000198/1].
The authors gratefully acknowledge the Gauss Centre for Supercomputing e.V. (www.gauss-centre.eu) for funding this project by providing computing time on the GCS Supercomputer JUWELS at J{\"u}lich Supercomputing Centre (JSC).
This work was performed using the Cambridge Service for Data Driven Discovery (CSD3), part of which is operated by the University of Cambridge Research Computing on behalf of the STFC DiRAC HPC Facility (www.dirac.ac.uk). The DiRAC component of CSD3 was funded by BEIS capital funding via STFC capital grants ST/P002307/1 and ST/R002452/1 and STFC operations grant ST/R00689X/1. DiRAC is part of the National e-Infrastructure.
This research was undertaken with the assistance of resources from the National Computational Infrastructure (NCI Australia), an NCRIS enabled capability supported by the Australian Government.
NumPy and SciPy
\citep{oliphant2007a}, IPython \citep{perez2007a}, Matplotlib
\citep{hunter2007a} and \href{https://github.com/artis-mcrt/artistools}{\textsc{artistools}}\footnotemark
\footnotetext{https://github.com/artis-mcrt/artistools} were used for data processing and plotting.

\section*{Data Availability}

The angle-dependent spectra are available on \href{https://doi.org/10.5281/zenodo.7997388}{Zenodo}\footnotemark
\footnotetext{https://doi.org/10.5281/zenodo.7997388}






\bibliographystyle{mnras}
\bibliography{astrofritz} 








\bsp	
\label{lastpage}
\end{document}